\documentclass[letterpaper,english,twocolumn,showpacs,preprintnumbers,amsmath,amssymb]{revtex4}
\usepackage[T1]{fontenc}
\usepackage{prettyref}
\usepackage{amsmath}
\usepackage{graphicx}

\makeatletter

\providecommand{\tabularnewline}{\\}

\usepackage{graphicx}
\usepackage{dcolumn}
\usepackage{bm}
\usepackage{multirow}

\def\bec{\begin{center}}
\def\eec{\end{center}}
\def\beq{\begin{equation}}
\def\eeq{\end{equation}}
\def\benu{\begin{enumerate}}
\def\eenu{\end{enumerate}}
\def\bit{\begin{itemize}}
\def\eit{\end{itemize}}
\def\beqr{\begin{eqnarray}}
\def\eeqr{\end{eqnarray}}
\def\beqrs{\begin{eqnarray*}}
\def\eeqrs{\end{eqnarray*}}
\def\btab{\begin{tabbing}}
\def\etab{\end{tabbing}}
\def\btable{\begin{tabular}}
\def\etable{\end{tabular}}

\@ifundefined{showcaptionsetup}{}{%
 \PassOptionsToPackage{caption=false}{subfig}}
\usepackage{subfig}
\makeatother

\usepackage{babel}

\begin{document}

\preprint{submitted to HB2008 PRST-AB Special Edition}

\title{Simulations of beam-beam and beam-wire interactions in RHIC}

\author{Hyung J. Kim }

\email{hjkim@fnal.gov}

\homepage{http://www-ap.fnal.gov/~hjkim}

\author{Tanaji Sen}

\affiliation{Fermi National Accelerator Laboratory, Batavia, Illinois 60510, USA}

\author{Natalia P. Abreu and Wolfram Fischer}

\affiliation{Brookheaven National Laboratory, Upton, NY 11973, USA}
\begin{abstract}
The beam-beam interaction is one of the dominant sources of emittance
growth and luminosity lifetime deterioration. A current carrying wire
has been proposed to compensate long-range beam-beam effects in the
LHC and strong localized long-range beam-beam effects are experimentally
investigated in the RHIC collider. Tune shift, beam transfer function,
and beam loss rate are measured in dedicated experiments. In this
paper, we report on simulations to study the effect of beam-wire interactions
based on diffusive apertures, beam loss rates, and beam transfer function
using a parallelized weak-strong beam simulation code (\texttt{bbsimc}).
The simulation results are compared with measurements performed in
RHIC during 2007 and 2008. 
\end{abstract}

\pacs{29.20.db, 29.27.Bd}

\keywords{wire compensator, beam-beam interaction}

\maketitle

\section{introduction}

Long-range beam-beam interactions are known to cause emittance growth
or beam loss in the Tevatron \cite{Sen,Shiltsev} and are expected
to deteriorate beam quality in the LHC. A possible remedy to reduce
their effects is to increase the crossing angle. However, it has a
side effect of reducing the luminosity due to the reduction in geometric
overlap. Compensation of long-range beam-beam interactions by applying
external electromagnetic forces has been proposed for the LHC \cite{Koutchouk}.
At large beam-beam separation, the electromagnetic force which a beam
exerts on individual particles of the other beam is proportional to
$\frac{1}{r}$, which can be generated and canceled out by the magnetic
field of a current-carrying wire. Direct-current wires were installed
in SPS \cite{Koutchouk-2}, DA$\Phi$NE \cite{Milardi-1}, and later
in RHIC \cite{Fischer-2} for tests. Results of the SPS wire excitation
experiments have been reported earlier \cite{Zimmermann,Dorda}. During
the KLOE experiment in DA$\Phi$NE, an improvement of beam lifetime
without luminosity loss due to the compensation of parasitic collisions
has been observed \cite{Milardi-1}. Experiments in RHIC supplement
the SPS and DA$\Phi$NE tests. The beam lifetime in RHIC is typical
for a collider storage ring and better than in the other two machines.
Unlike the SPS, two beams circulate and interact in RHIC and compensation
of a single long-range interaction is possible. The energies and other
beam conditions are also different from the SPS and DA$\Phi$NE. So
the results reported here will result in simulations being benchmarked
under a wider range of operating conditions.

Two current carrying wires, one for each beam, have been installed
between the magnets $Q3$ and $Q4$ of IP6 in the RHIC tunnel. Their
impact on a beam was measured during the physics run with deuteron
and gold beams. No attempt was made to compensate the beam-beam interaction
since so far only ion beams were available which have a beam-beam
parameter about a factor of 3 smaller than proton beams. However,
the experimental results help to understand the beam-beam effects
because the wire force is similar to the long-range beam-beam force
at large separations. In this report we discuss the results of numerical
simulations of a wire acting on a beam in RHIC using a multi-particle
tracking code and compare with measurements in RHIC Run 7 and 8 \cite{apex}. 

The organization of the paper is as follows: The physical models used
in the simulation code are described in Section \ref{sec:Model}.
The beam and wire parameters are given in Section \ref{sec:parameters}.
Section \ref{sec:results} presents the simulation results of the
effects of the wire on the beam dynamics and the comparison between
measurement and simulation results. We summarize our results in Section
\ref{sec:summary}.

\section{Model \label{sec:Model}}

In the collider simulation, the two beams moving in opposite direction
are represented by macroparticles of which the charge to mass ratio
is that of each beam. The number of macroparticles is much less than
the bunch intensity of the beam because even with modern supercomputers
it is too time consuming to track 10$^{11}$ particles for the number
of revolutions of interest. The macroparticles are generated and loaded
with an initial distribution for a specific simulation purpose, for
example, six-dimensional Gaussian distribution for long-term beam
evolution. The transverse and longitudinal motion of particles is
calculated by transfer maps which consist of linear and nonlinear
maps. In the simulations, the following nonlinearities are included:
head-on and long-range beam-beam interactions, external electromagnetic
force by current carrying wire, multipole errors in the interaction
region (IR) quadrupole triplets, and sextupoles for chromaticity correction.
The finite bunch length effect of the beam-beam interactions is considered
by slicing the beam into several sections in the longitudinal direction
and by a synchro-beam map \cite{Hirata}. Each slice in a beam interacts
with particles in the other beam in turn at the collision points.
In addition, the magnetic field of a finite length of wire is calculated
using the Biot-Savart law at the particles' betatron amplitude.

The six-dimensional accelerator coordinates $\mathbf{x}=\left(x,x^{'},y,y^{'},z,\delta\right)^{T}$
are applied, where $x$ and $y$ are horizontal and vertical coordinates,
$x^{'}$ and $y^{'}$ the trajectory slopes of each coordinates, $z=-c\Delta t$
the longitudinal distance from the synchronous particle, and $\delta=\Delta p_{z}/p_{0}$
the momentum deviation from the synchronous energy. Coupling between
the transverse planes is included in the transfer map between elements.
We adopt the weak-strong model to treat the beam-beam interactions
where one beam is strong and is not affected by the other beam while
the other beam is weak and experiences a beam-beam force due to the
strong beam during the interactions. The density distribution of the
strong beam is assumed to be Gaussian.

It is well known that for a large separation distance $\left(\gg\sigma\right)$
at parasitic crossings, the strength of long-range interactions is
inversely proportional to the distance. Its effect on a test beam
can be compensated by current carrying wires which create just the
$\frac{1}{r}$ field. The advantage of such an approach consists of
the simplicity of the method and the possibility to deal with all
multipole orders at once. For a finite length of a wire embedded in
the middle of a drift length $L$ and tilted in pitch and yaw angles,
the transfer map of a wire can be written as \cite{Erdelyi-1} \begin{equation}
\begin{aligned}\mathcal{M}_{w} & =S_{\Delta x,\Delta y}\odot T_{\theta_{x},\theta_{y}}^{-1}\odot D_{L/2}\odot\mathcal{M}_{k}\\
 & \quad\odot D_{L/2}\odot T_{\theta_{x},\theta_{y}},\end{aligned}
\label{eq:wc-1}\end{equation}
 where $T_{\theta_{x},\theta_{y}}$ represents the tilt of the coordinate
system by horizontal and vertical angles $\theta_{x},\theta_{y}$
to orient the coordinate system parallel to the wire, $D_{L/2}$ is
the drift map with a length $\frac{L}{2}$, $\mathcal{M}_{k}$ is
the wire kick integrated over a drift length, and $S_{\Delta x,\Delta y}$
represents a shift of the coordinate axes to make the coordinate systems
after and before the wire agree.

At the start of the simulation, the particles in the weak beam are
distributed over the phase space $\mathbf{x}=\left(x,x^{'},y,y^{'},z,\delta\right)^{T}$.
The number of particles $N$ is limited by the computational power.
In order to make the best use of a relatively small number of simulation
particles compared to the bunch intensity, the initial distribution
should be optimized. Indeed the initial distribution is very important
because a proper choice can reduce the statistical noise in the physical
quantities. The simulation particles are generated in two steps:
\begin{enumerate}
\item The particle coordinates $\left(x,y,z\right)$ of particles can be
directly generated from the spatial Gaussian distribution, $\bar{\rho}\left(x,y,z\right)=\bar{\rho}_{x}\left(x\right)\bar{\rho}_{y}\left(y\right)\bar{\rho}_{z}\left(z\right)$,
where $\bar{\rho}_{\zeta}\left(\zeta\right)=\bar{\rho}_{\zeta0}\exp\left(-\frac{\zeta^{2}}{2\sigma_{\zeta}^{2}}\right)$.
Since the particle coordinates are not correlated, one can generate
them by inverse mapping of each cumulative distribution function of
horizontal, vertical, and longitudinal Gaussian distributions, and
together with the bit-reversed sequence to minimize nonphysical correlations
\cite{Birdsall}.
\item An equilibrium distribution in transverse phase space e.g. in the
$(x,x')$ plane is $\hat{\rho}\left(x,x^{'}\right)=\hat{\rho}_{0}\exp\left(-\frac{x^{2}+\left(\beta_{x}x^{'}+\alpha_{x}x\right)^{2}}{2\sigma}\right)$.
Since the spatial coordinate $x$ is determined at the first step,
the slope $x^{'}$ can be obtained from the random variate $r$ of
a univariable Gaussian, i.e., $x^{'}=\left(r-\alpha_{x}x\right)/\beta_{x}$. 
\end{enumerate}
Following the above physical model, a beam-beam simulation code \texttt{bbsimc}
has been developed at FNAL over the past few years to study the effects
of the machine nonlinearities and the beam-beam interactions \cite{bbsimc}.
If required, time dependent effects such as tune modulation and fluctuation,
beam offset modulation and fluctuation, dipole strength fluctuations
to mimic rest-gas scattering etc can be included in the model. The
code is under continuous development with the emphasis being on including
the important details of an accelerator and the ability to reproduce
observations in diagnostic devices. At present, the code can be used
to calculate tune footprints, dynamic apertures, beam transfer functions,
frequency diffusion maps, action diffusion coefficients, emittance
growth and beam lifetime. Calculation of the last two quantities over
the long time scales of interest is time consuming even with modern
computer technology. In order to run efficiently on a multiprocessor
system, the resulting model was implemented by using parallel libraries
which are MPI (interprocessor Message Passing Interface standard)
\cite{mpi}, state-of-the-art parallel solver libraries (Portable,
Extensible Toolkit for Scientific Calculation, PETSc) \cite{petsc},
and HDF5 (Hierarchical Data Format) \cite{hdf5}.

\section{Wire and Beam Parameters\label{sec:parameters}}

The transverse electric field of a round Gaussian bunch with transverse
rms size $\sigma$ is \begin{align}
\vec{E} & =\frac{n_{*}q_{*}}{2\pi\epsilon_{0}}\frac{\vec{r}}{r^{2}}\left(1-\exp\left(-\frac{r^{2}}{2\sigma^{2}}\right)\right),\label{eq:rd-1}\end{align}
 where $n_{*}$ and $q_{*}$ are the number of particles and the electric
charge in the opposite bunch respectively, and $r$ is the radial
distance from the center of the beam. At a small radius $r\ll\sigma$,
the field is proportional to $r$ and shifts the betatron tune. This
tune shift is characterized by the beam-beam parameter $\xi=\frac{n_{*}r_{0}}{4\pi\epsilon}$,
where $r_{0}$ is the classical radius of the particle, $\epsilon=\gamma\sigma^{2}/\beta^{2}$
is the normalized emittance. While particles at small $r$ undergo
a linear tune shift, the particles with $r\gg\sigma$ experience a
$\frac{1}{r}$ force. The long-range effect is nonlinear and may vary
from bunch to bunch if the the bunch pattern is asymmetric. A current
carrying wire generates a magnetic force which is $\propto\frac{1}{r}$,
the same as the long-range beam-beam force at large separations. The
wire current required to compensate a long-range interaction is $\left(I_{w}L_{w}\right)=n_{*}q_{*}c$,
where $I_{w}$ is the wire current, and $L_{w}$ its length. In addition
to the wire strength, the phase advance from the location of the wire
to the location of the long-range interaction should be small for
an effective compensation of the long-range interaction \cite{Zimmermann-2}.
In RHIC, a possible parasitic collision point is near the separation
dipole DX magnets near IP6 and IP8. There is sufficient space in the
straight sections between quadrupoles $Q_{3}$ and $Q_{4}$ for the
wires. The phase advance between the DX magnet and the wire location
is 5.7$^{\circ}$ for $\beta^{*}=$ 1 m and is expected to provide
effective compensation \cite{Fischer-1}.

\begin{figure}
\begin{centering}
\subfloat[]{

\includegraphics[scale=0.75]{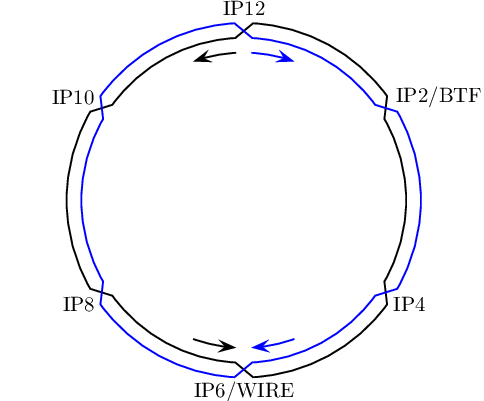}}
\par\end{centering}

\begin{centering}
\subfloat[]{

\includegraphics{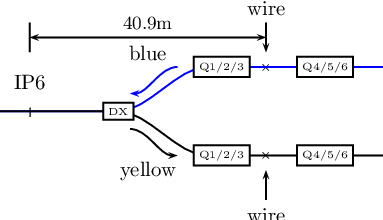}}
\par\end{centering}

\caption{(a) Layout of RHIC, and (b) location of the long-range beam-beam interaction
and wire compensators at IP6. Beam-beam collisions occur at IP6 (STAR)
and IP8 (PHOENIX).\label{fig:layout}}

\end{figure}

Figure \ref{fig:layout} (a) shows the overall layout of RHIC which
has two head-on beam-beam collisions at IP6 and IP8. %
 The Yellow beam revolves in a counter clockwise direction, and the
Blue beam revolves clockwise. The elements and the location of the
wire at IP6 are shown in Fig. \ref{fig:layout} (b). Two wires are
currently installed in RHIC: one for each beam. In these simulations
the effects of a current carrying wire on a beam have been studied
for different RHIC runs: gold beam at both injection and store energies,
and deuteron beam at store energy. The parameters of RHIC used in
our simulations are shown in Table \ref{tab:rhic}. %
\begin{table*}
\begin{centering}
\begin{tabular}{ccccc}
\hline 
\multirow{2}{1.cm}{quantity} & \multirow{2}{0.5cm}{unit} & \multicolumn{2}{c}{gold beam} & \multirow{2}{2.5cm}{deuteron beam}\tabularnewline
 &  & injection & store & \tabularnewline
\hline
energy & Gev/nucleon & 9.795 & 100 & 107.396 \tabularnewline
bunch intensity (Blue/Yellow) & $10^{9}$ & 0.7/0.7 & 1/1 & 134/1\tabularnewline
emittance $\epsilon_{x,y}$(95\%) & mm mrad & 5.8 & 18 & 17\tabularnewline
$\left(\beta_{x}^{*},\beta_{y}^{*}\right)$ at IP6 & m & $\left(9.93,10.0\right)$ & $\left(0.97,0.94\right)$ & $\left(0.87,0.89\right)$\tabularnewline
$\left(\beta_{x},\beta_{y}\right)$ at wire location & m & $\left(119.7,34.6\right)$ & $\left(1082,392\right)$ & $\left(1194,393\right)$\tabularnewline
beam-beam parameter $\xi$ & $10^{-3}$ &  & 1.25 & 1.46\tabularnewline
revolution frequency & kHz & 77.8 & 78.2 & 78.2\tabularnewline
tunes $\left(\nu_{x},\nu_{y}\right)$: Blue &  & $\left(0.230,0.216\right)$ & $\left(0.220,0.231\right)$ & $\left(0.235,0.225\right)$\tabularnewline
tunes $\left(\nu_{x},\nu_{y}\right)$: Yellow &  & $\left(0.220,0.230\right)$ & $\left(0.232,0.228\right)$ & $\left(0.225,0.235\right)$\tabularnewline
chromaticity $\left(\xi_{x},\xi_{y}\right)$ &  & $\left(2,2\right)$ & $\left(2,2\right)$ & $\left(2,2\right)$\tabularnewline
\hline
\textsf{$\left(IL\right)_{max}$}  & Am &  & \textsf{125}  & \tabularnewline
\textsf{$L_{w}$}  & m &  & \textsf{2.5}  & \tabularnewline
\textsf{$r_{w}$}  & mm &  & \textsf{3.5}  & \tabularnewline
vertical separation $\left(d_{y}\right)$ & $\sigma$ & 7-14 & 6-10 & 7-12\tabularnewline
\hline
\end{tabular}
\par\end{centering}

\caption{RHIC parameters for gold beam at injection and store, and deuteron
beam at store. \label{tab:rhic}}

\end{table*}
 The length of each wire is 2.5 m. The wire current can be varied
to a maximum of 50 A. Each wire can be moved in the vertical direction
over a range of 65 mm.

\section{Effects of a wire on a beam \label{sec:results}}

The integrated magnetic fields $\vec{B}$ from a wire are found from
\cite{Erdelyi-1} \begin{align}
\left(\begin{array}{c}
B_{x}\\
B_{y}\end{array}\right) & =\frac{\mu_{0}I_{w}}{4\pi}\frac{u-v}{x^{2}+y^{2}}\left(\begin{array}{c}
x\\
y\end{array}\right),\label{eq:rd-6}\end{align}
 where $I_{w}$ is the current of wire compensator, $u$ and $v$
are $\sqrt{\left(\frac{3L_{w}}{2}\right)^{2}+x^{2}+y^{2}}$ and $\sqrt{\left(\frac{L_{w}}{2}\right)^{2}+x^{2}+y^{2}}$
respectively, and $L_{w}$ the length of the wire. For the wire kick,
the the field and the force exerted on the beam are $\propto1/r$
where $r$ is the transverse distance from the beam to the wire. This
force is nonlinear and is expected to have a significant impact on
the beam quality at sufficiently close distances.

The impact of a wire can be observed in RHIC by measuring the orbit
change, tune shift, the beam transfer function and the loss rates.
The tune shift is one of the fundamental observables and it can be
directly verified with analytical calculation. However, numerical
simulations allow us to calculate other quantities not easily observable
but which give valuable insight into the beam dynamics and can complement
the experiments. These numerically calculable quantities include the
tune footprint, the frequency diffusion map, the dynamic aperture,
and the diffusion coefficients to characterize the diffusion in action.
In this section, we present the simulation results of all the above
quantities at RHIC injection and collision runs, and compare them
to measurements where possible.

In the model for dynamics at injection energy we include the chromaticity
sextupoles, these being the dominant nonlinearity, and the current
carrying wire. It should be noted that the persistent current sextupoles
in the arc dipoles were not included in the model because their inclusion
would have increased the computation time substantially and they have
a smaller impact on the dynamics compared to the chromaticity correcting
sextupoles. At collision energy, the beam is strongly influenced by
both machine nonlinearities and beam-beam interactions. In the collision
model, we include the chromaticity sextupoles, the nonlinear field
errors in the IR triplets, and the head-on beam-beam interactions
at IP6 and IP8, together with the wire. There are two variables associated
with the wire that can be varied: the wire current and the beam-wire
separation. In the simulations the wire strength is set to the maximum
value of 125 Am in the RHIC wire compensator so that the effects due
to the wire may be clearly evident. These results will be compared
with the data taken at this current during the machine experiments
of 2007 and 2008. Depending on the purpose of the simulation, the
beam-wire separation is also varied typically in the range of 4-10
$\sigma$ where $\sigma$ is the rms beam size at the wire location.
The beam-wire separation is only in the vertical direction. The Blue
wire is placed below, the Yellow wire above the beam. The simulations
reported here will be for the Blue beam.

\subsection{Tune shift and tune footprint}

A basic check of the simulation is to compare the simulated tune shifts
with the measured tune shifts and also against analytical calculation.
The transverse tune shift at zero amplitude due to wire kicks is given
by \begin{align}
\Delta\nu_{x,y} & =\pm\frac{\mu_{0}I_{w}L_{w}}{8\pi^{2}\left(B\rho\right)\sigma^{2}}\beta_{x,y}\frac{d_{y}^{2}-d_{x}^{2}}{\left(d_{y}^{2}+d_{x}^{2}\right)^{2}},\label{eq:rd-2}\end{align}
 where $d_{x,y}$ denote the beam-wire distances normalized by $\sigma_{x,y}$,
the rms beam sizes at the wire location, $I_{w}L_{w}$ the integrated
strength of the wire, and $\beta_{x,y}$ the beta functions at wire
location. At the wire location, the ratio of horizontal to vertical
beta functions is $\beta_{x}/\beta_{y}=2.8$ for the gold beam at
store, and $\beta_{x}/\beta_{y}=3.0$ for the deuteron beam. For $I_{w}L_{w}=125$
Am which is the maximum integrated strength allowed in the RHIC wire
installation, we get the horizontal tune shifts at 7 $\sigma$ vertical
separation: $\Delta\nu_{x}=5.1\times10^{-3}$ for the gold beam, and
$\Delta\nu_{x}=5.9\times10^{-3}$ for the deuteron beam at store.
Without the wire, beam-beam collisions are a major source of tune
spread and shift. The beam-beam parameters for two head-on collisions
are $\xi=2.4\times10^{-3}$ and $\xi=2.8\times10^{-3}$ for gold and
deuteron beams at store energy respectively. These tune shifts are
calculated for zero amplitude particles. For beam-wire separations
greater than 6 $\sigma$, the large amplitude particles experience
a detuning and shift mainly from the wire while the tune shift of
small amplitude particles is dominated by the head-on collisions.

\begin{figure}
\begin{centering}
\subfloat[]{

\centering{}\includegraphics[bb=140bp 255bp 465bp 525bp,clip,scale=0.5]{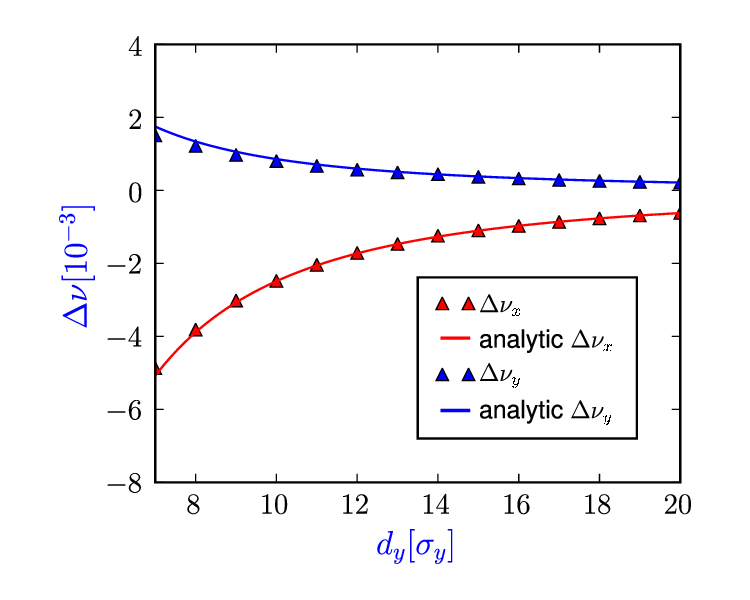}}
\par\end{centering}

\begin{centering}
\subfloat[]{

\centering{}\includegraphics[scale=0.5]{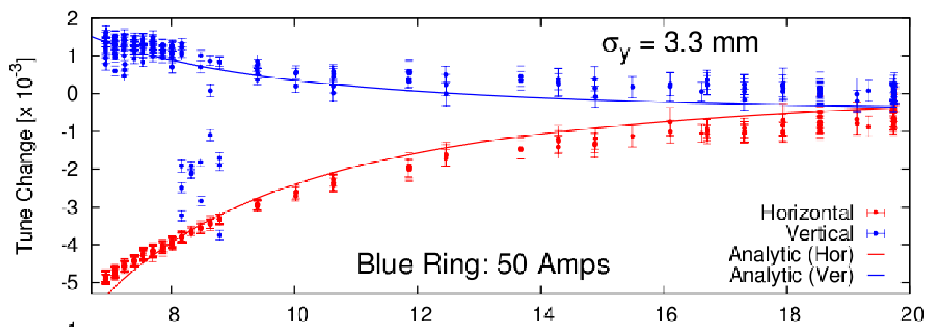}}
\par\end{centering}

\caption{Plots of tune shift dependence on the wire separation distance: (a)
simulation and (b) measurement \cite{Fischer-2}. Data set are obtained
at gold beam at store energy.\label{fig:tune-shift}}

\end{figure}

In our simulations, particle tunes are calculated with a Hanning filter
applied to an FFT of particle coordinates found from tracking. Figure
\ref{fig:tune-shift} (a) compares the analytical expression in Eq.
\ref{eq:rd-2} with the tune shift from tracking and averaged over
a Gaussian distribution of particles. The changes in the transverse
tune as a function of the beam-wire vertical separation distance in
units of $\sigma_{y}$ is in good agreement with the analytic relation
Eq. \ref{eq:rd-2}. The relative difference in vertical tune shift
between the simulation and theory is $\approx10\%$ at $d_{y}=$ 7
$\sigma$, which stems from our usage of the centroid to calculate
the tune shift and not the zero amplitude particle. The measured tune
shifts in the recent experiments are compared with the analytical
values in Fig. \ref{fig:tune-shift}(b). These show that measurements
and simulations also agree. %

\begin{figure}
\begin{centering}
\subfloat[]{

\centering{}\includegraphics[bb=140bp 255bp 470bp 525bp,clip,scale=0.5]{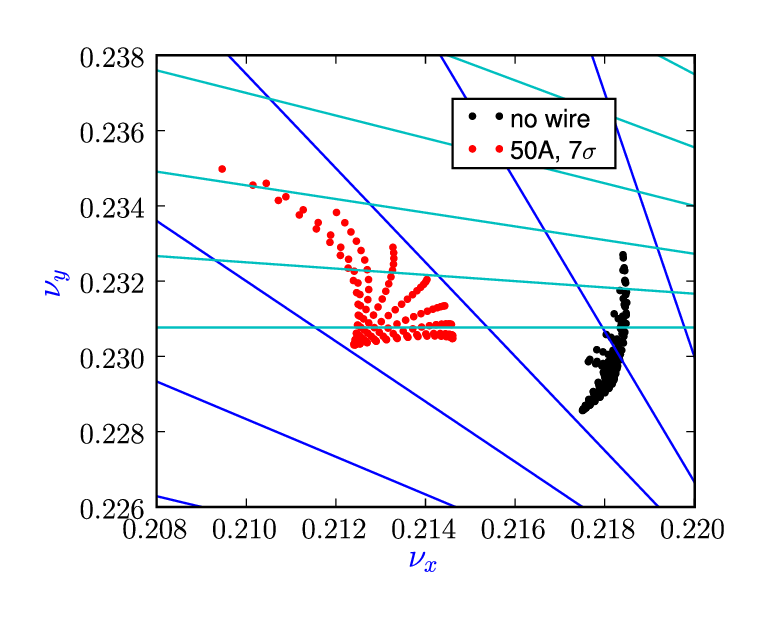}}
\par\end{centering}

\begin{centering}
\subfloat[]{

\centering{}\includegraphics[bb=140bp 255bp 470bp 525bp,clip,scale=0.5]{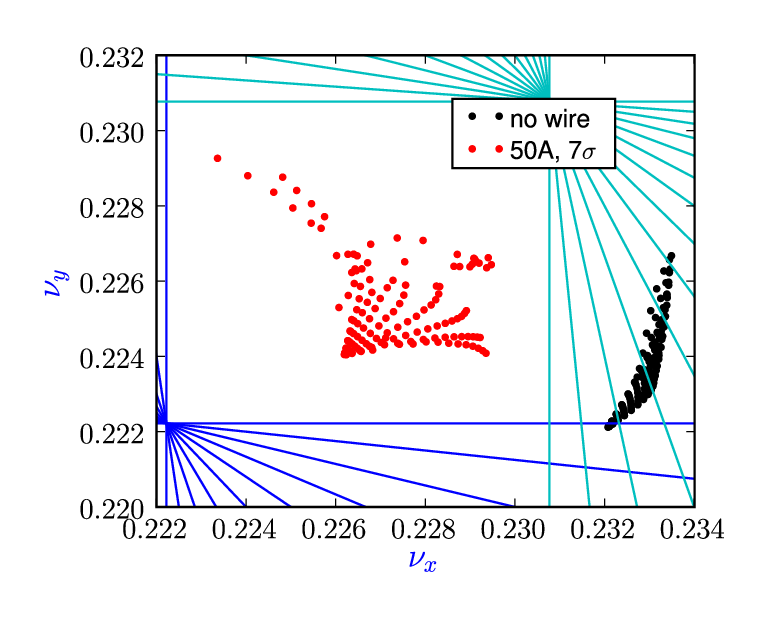}}
\par\end{centering}

\caption{Plots of tune footprints for (a) gold beam and (b) deuteron beam at
store energy. The no-wire case includes sextupoles, IR multipoles,
and head-on collisions. The wire current is 50A and wire-beam separation
distance sets to 7$\sigma$. Blue and cyan lines stand for 9$^{th}$
and 12$^{th}$ order resonances respectively.\label{fig:footprint}}

\end{figure}

The tune footprint also provides useful information especially on
the choice of working point and in finding the resonances spanned
by the beam distribution. Figure \ref{fig:footprint} shows tune footprints
from tracking single particles with initial amplitudes in the range
0-4 $\sigma_{x,y}$ for two cases: without the wire and with the wire
powered at 125 Am and 7$\sigma$ vertical separation from the beam.
Comparing the two footprints without and with the wire, we observe
that the horizontal tune shift due to the wire is different from the
vertical one. The ratio of horizontal to vertical tune shifts is,
for example, $\Delta\nu_{x}/\Delta\nu_{y}=2.8$ for the gold beam,
as shown in Fig. \ref{fig:footprint} (a), which agrees well with
the ratio of beta functions at the wire. We observe that the width
of the footprint is wider at large amplitude due to its dependence
on the separation distance between a target particle and the wire.
The wire alone causes the horizontal tune spread, for example, at
4 $\sigma$ amplitude to be 10 times larger than for the no wire case.
Characteristics of tune footprints of the deuteron beam, seen in Fig.
\ref{fig:footprint} (b), are similar to that of the gold beam. %
 However, the final tunes of the deuteron beam due to the wire is
closer to the diagonal than the gold beam tunes due to the difference
of their nominal tunes. Both the $9^{th}$ and $12^{th}$ order resonances
are spanned by the gold beam while the deuteron beam is free from
these resonances. The tune shifts are slightly larger for the deuteron
beam than the  gold beam because of the difference of beta functions
and beam sizes.

\subsection{Frequency diffusion}

We have calculated frequency diffusion maps as another way to investigate
the effects of a current carrying wire. The map represents the variation
of the betatron tunes over two successive sets of the tunes \cite{Laskar}:
The variation can be quantified by $d=\log\sqrt{\Delta\nu_{x}^{2}+\Delta\nu_{y}^{2}}$,
where $\Delta\nu_{x}=\nu_{x}^{\left(2\right)}-\nu_{x}^{\left(1\right)}$
is the horizontal tune variation, in the simulations, between the
first set of 1024 turns and the next set of 1024 turns, and $\Delta\nu_{y}=\nu_{y}^{\left(2\right)}-\nu_{y}^{\left(1\right)}$.
If the tunes $\left(\nu_{x}^{\left(1\right)},\nu_{y}^{\left(1\right)}\right)$
are different from $\left(\nu_{x}^{\left(2\right)},\nu_{y}^{\left(2\right)}\right)$,
the particle's orbit diffuses. A large tune variation is generally
an indicator of reduced stability. 

\begin{figure}
\begin{centering}
\subfloat[]{

\centering{}\includegraphics[bb=145bp 250bp 485bp 530bp,clip,scale=0.5]{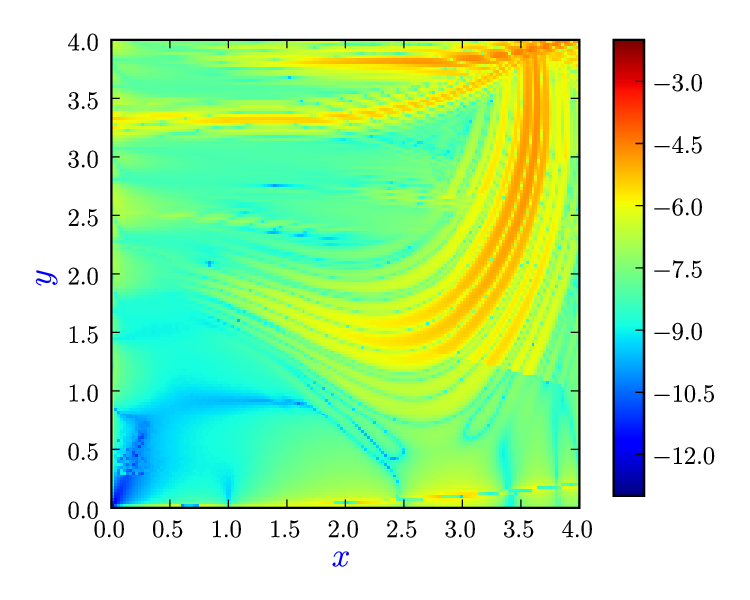}}
\par\end{centering}

\begin{centering}
\subfloat[]{

\centering{}\includegraphics[bb=145bp 250bp 485bp 530bp,clip,scale=0.5]{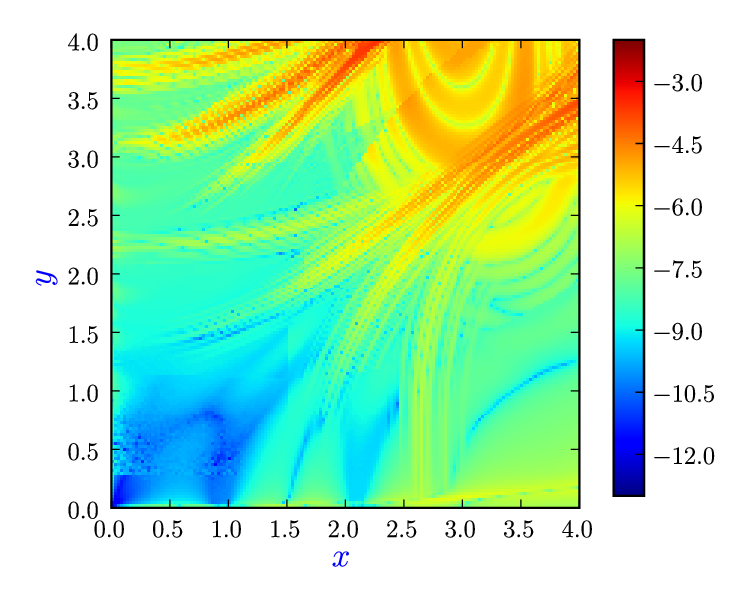}}
\par\end{centering}

\caption{Plot of frequency diffusion map of betatron tunes (a) without and
(b) with wire for gold beam at collision energy. Wire strength is
125 Am, and wire-beam separation is 7$\sigma$. The tune change is
logarithmically scaled by $\log\sqrt{\Delta\nu_{x}^{2}+\Delta\nu_{y}^{2}}$.
\label{fig:fd_au}}

\end{figure}

\begin{figure}
\begin{centering}
\subfloat[]{

\centering{}\includegraphics[bb=145bp 250bp 485bp 530bp,clip,scale=0.5]{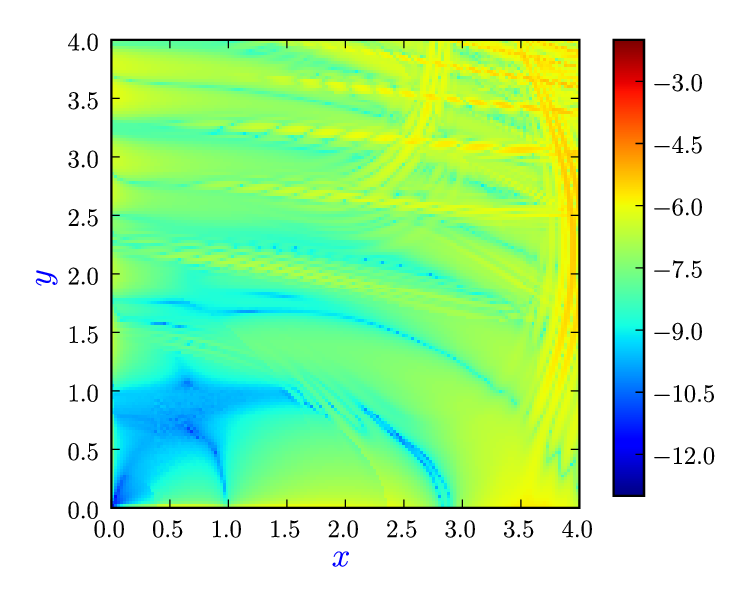}}
\par\end{centering}

\begin{centering}
\subfloat[]{

\centering{}\includegraphics[bb=145bp 250bp 485bp 530bp,clip,scale=0.5]{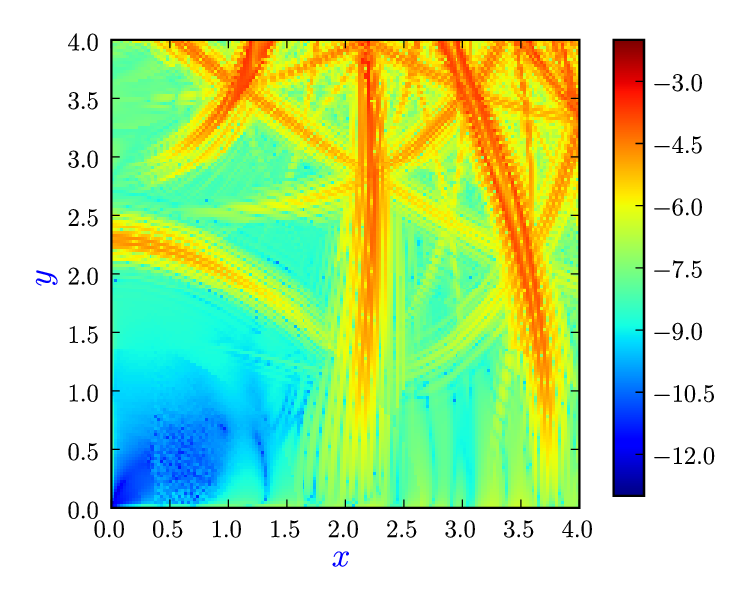}}
\par\end{centering}

\caption{Plot of frequency diffusion map of betatron tunes (a) without and
(b) with wire for deuteron beam at collision energy. Wire strength
is 125 Am, and wire-beam separation is 7 $\sigma$. The tune change
is logarithmically scaled by $\log\sqrt{\Delta\nu_{x}^{2}+\Delta\nu_{y}^{2}}$.
\label{fig:fd_dau}}

\end{figure}

Figure \ref{fig:fd_au} shows the frequency diffusion map of the betatron
tune for the gold beam at collision energy and the influence of the
wire on the map. %
 The color scale shown on the right of each map gives a quantitative
measure of the diffusion: red color corresponds to larger diffusion,
and blue color represents less diffusion. In Fig. \ref{fig:fd_au}
we directly see the lines connected to resonances. The wire increases
the detuning of betatron tune and makes the particle motions more
chaotic at amplitude beyond 3 $\sigma$. The stability boundary is
shrunk further. The diffusion map, Fig. \ref{fig:fd_dau} (a), of
the deuteron beam without the wire shows mostly stable motion and
only a small region with appreciable diffusion. It doesn't show the
traces of resonances and looks quite different from that of the gold
beam. %
 This is to be expected since the resonances spanning the footprint
of the deuteron beam are only 12$^{th}$ order. However, the wire
changes the diffusion map significantly, as is evident in Fig. \ref{fig:fd_dau}
(b). While the region of high diffusion for the gold beam is distributed
at large amplitudes, the regions with large diffusion are observed
even at small amplitudes with the  deuteron beam. The red {}``tongues''
even extend to particles in the core, e.g. to particles at 0.5 $\sigma$.
This is to be contrasted with the tune footprint in Fig \ref{fig:footprint}
which shows that no resonances below 12th order are spanned by the
beam distribution. In later sections we will discuss the correlation
of these frequency diffusion maps with the dynamic aperture and loss
rates.

\subsection{Dynamic Aperture\label{sub:DA}}

Magnet nonlinearities and beam-beam interactions limit the dynamic
aperture. The dynamic aperture quantifies detrimental effects of nonlinearities
and can be calculated relatively quickly. The dynamic aperture of
an accelerator is defined as the largest radial amplitude of particles
that survive up to a certain time interval; in this simulation, we
choose $10^{6}$ turns for both injection and collision which corresponds
to 13 seconds circulation time in RHIC. It is, for example, only about
10\% of the RHIC injection period for ions. We examined the dependence
of the aperture on the number of turns by calculating it for $10^{4}-10^{7}$
turns and found that the dynamic aperture stays nearly constant after
$10^{6}$ turns.

We have seen that a wire has a significant impact on the tune shift
and tune spread, for example, the tune shift due to 125 Am wire strength
is 0.005 at 7 $\sigma$ separation. The significant increase of the
frequency diffusion due to the wire implies that the motion of particles
becomes more chaotic at large amplitudes. Hence we can expect that
a wire will also have a significant impact on the dynamic aperture.

\begin{figure*}
\begin{centering}
\subfloat[]{

\centering{}\includegraphics[bb=140bp 255bp 465bp 525bp,clip,scale=0.5]{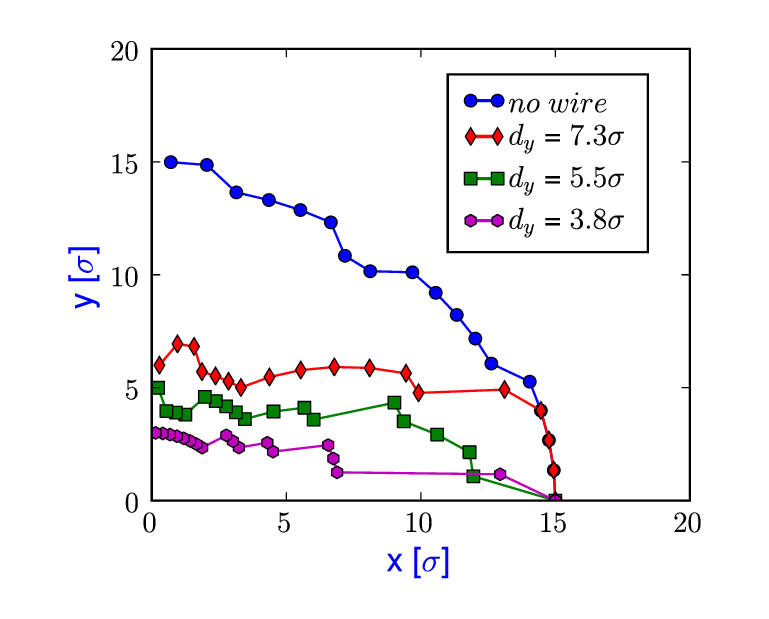}}\subfloat[]{

\centering{}\includegraphics[bb=140bp 255bp 465bp 525bp,clip,scale=0.5]{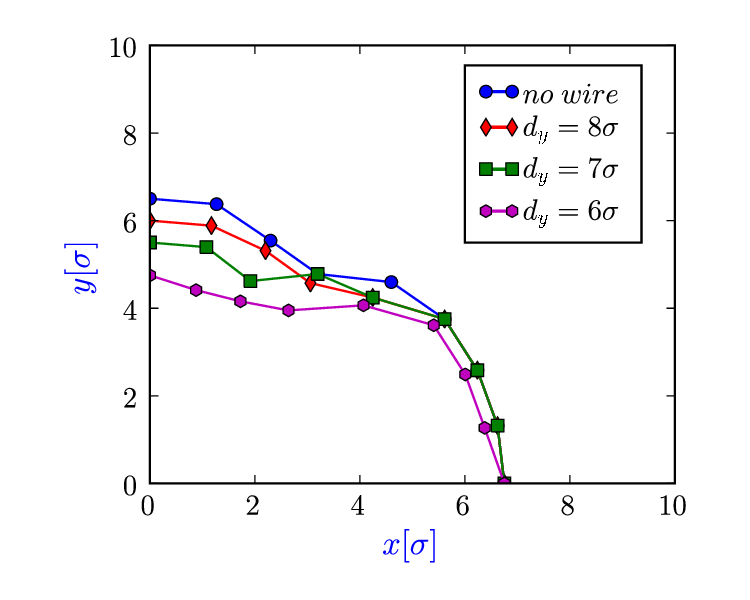}}\subfloat[]{

\centering{}\includegraphics[bb=140bp 255bp 465bp 525bp,clip,scale=0.5]{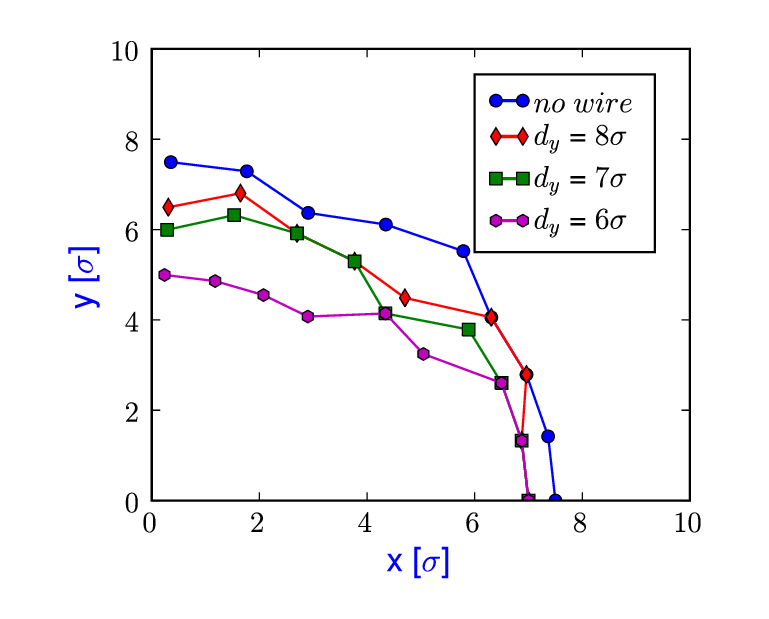}}
\par\end{centering}

\caption{Plots of dynamic aperture according to wire separation distance: (a)
gold beam injection energy, (b) gold beam collision energy, and (c)
deuteron beam collision energy. The wire strength is set 125 Am. \label{fig:dynamic}}

\end{figure*}

At injection energy the simulation model for dynamic aperture calculations
includes sextupoles, transverse coupling, as well as the nonlinearity
due to the wire. At collision energy, the nonlinearities due to the
multipoles in the IR quadrupoles are added and the head-on beam-beam
interactions. Particles are distributed uniformly over the transverse
planes with amplitudes 0-15 $\sigma$. The linear chromaticity is
set $\left(\xi_{x},\xi_{y}\right)=\left(+2,+2\right)$. Since non-zero
chromaticity increases the momentum dependent tune spread, the off-momentum
particles cross more betatron resonances during synchrotron oscillations.
The plots in Fig. \ref{fig:dynamic} show the dynamic aperture for
the gold beam at injection energy, the gold beam at collision energy
and the deuteron beam at collision energy respectively.%
 Plots (a) and (b) in Fig. \ref{fig:dynamic} show that nonlinearities
of the IR triplets play a major role in determining the stability
boundary. When the electromagnetic force due to the wire is not present,
the boundary at injection energy is at 15 $\sigma$ while this boundary
at collision energy falls to 7 $\sigma$ primarily due to the IR multipole
errors. The effect of the head-on beam-beam collisions on the dynamic
aperture is negligible over our simulation period since the head-on
interactions do not transport particles to large amplitudes. Sextupoles
have only a small impact at collision energy since their strengths
are relatively small compared to the IR multipole strengths. The stability
boundaries are approximately circular, but the wire distorts the boundaries
near the vertical plane since the beam-wire separation is entirely
in the vertical plane. The dynamic aperture near the vertical plane
decreases considerably as the wire approaches the beam. The stability
boundary along the vertical plane is directly proportional to the
wire separation while along the horizontal plane, the boundary is
nearly independent of the separation. The difference in the dynamic
aperture between the deuteron beam and the gold beam at collision
energies is approximately 1 $\sigma$ when the wire is not present.
This can be understood from the difference in resonances spanning
between both beams mentioned in the previous section. With the wire
powered, the dynamic aperture in the two cases is nearly the same.

We now turn to tune scans of the dynamic aperture. One of the key
parameters for machine operation is the working point. The search
for a working point with a good beam lifetime and high luminosity
is always a major issue. Since the dynamic aperture becomes smaller
when the betatron tune of beam particles is on or across resonances,
the tune diagram is often a useful indicator of single particle stability.
In the simulations, the tune scans are performed with increments of
$\Delta\nu=0.01$ in the two transverse directions. At each set of
the tune scan, we load identical distributions in order to avoid the
uncertainties due to different distribution and beam size reduction
from the previous tune.

\begin{figure}
\begin{centering}
\subfloat[]{

\centering{}\includegraphics[bb=135bp 250bp 470bp 530bp,clip,scale=0.5]{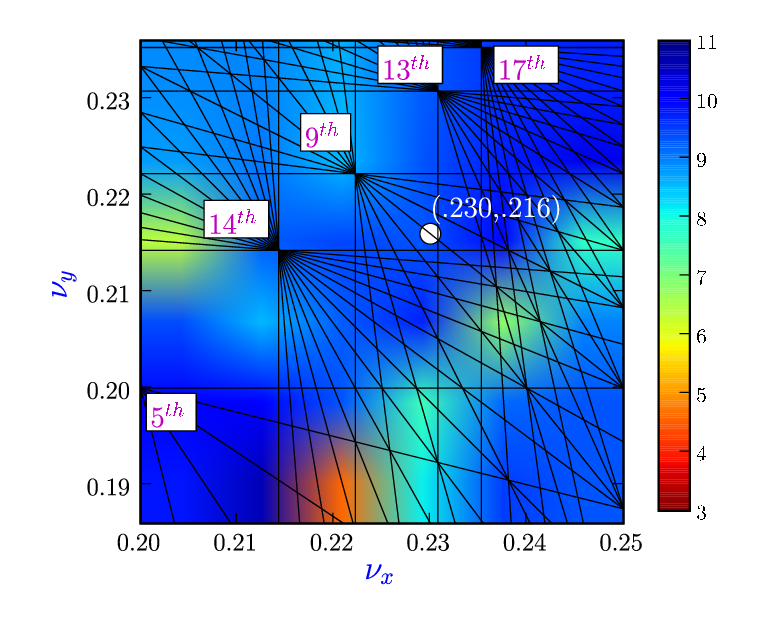}}
\par\end{centering}

\begin{centering}
\subfloat[]{

\centering{}\includegraphics[bb=135bp 250bp 470bp 530bp,clip,scale=0.5]{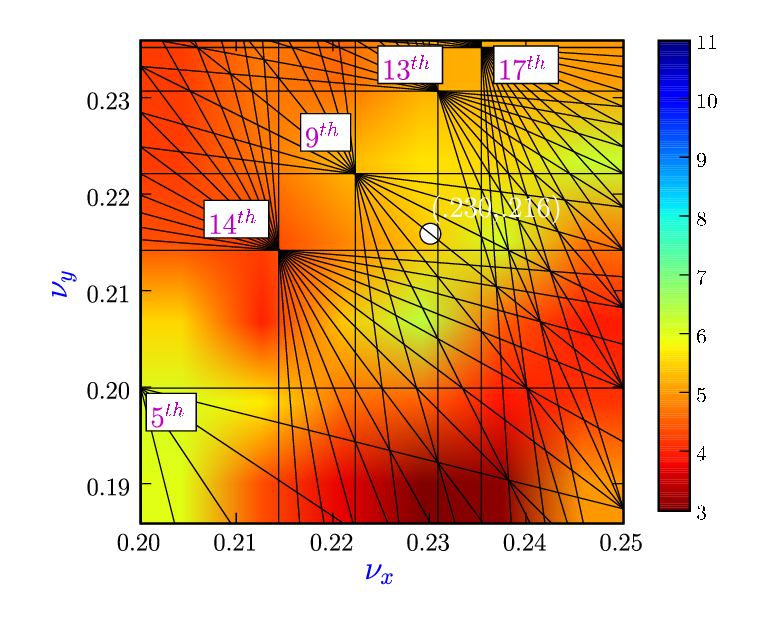}}
\par\end{centering}

\caption{Tune scan of dynamic aperture at gold injection energy: (a) $d_{y}=7.8\sigma$
and (b) $d_{y}=3.8\sigma$. Working point $\left(28.230,29.216\right)$
is plotted as a white dot. Red indicates low dynamic apertures while
blue indicates high dynamic apertures. The unit of dynamic aperture
is $\sigma$. \label{fig:da-ts-gi}}

\end{figure}

\begin{figure}
\begin{centering}
\subfloat[]{

\centering{}\includegraphics[bb=135bp 250bp 470bp 530bp,clip,scale=0.5]{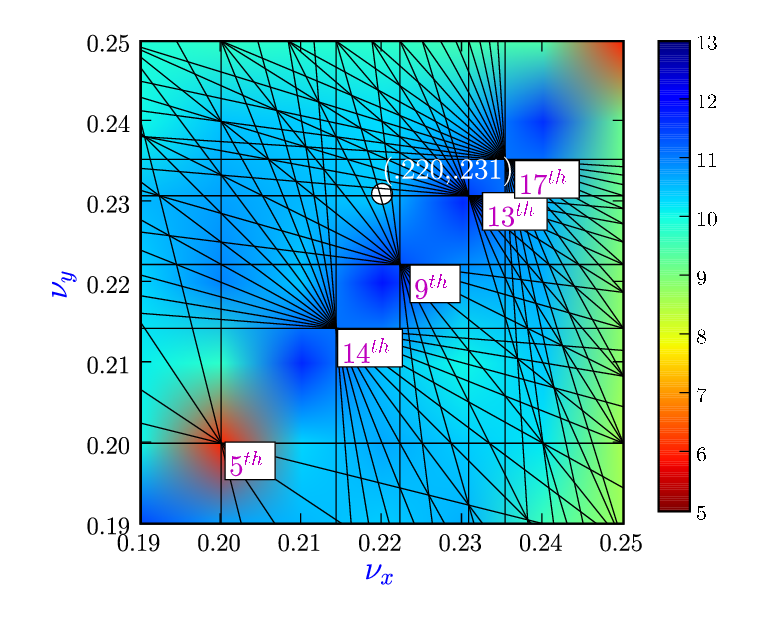}}
\par\end{centering}

\begin{centering}
\subfloat[]{

\centering{}\includegraphics[bb=135bp 250bp 470bp 530bp,clip,scale=0.5]{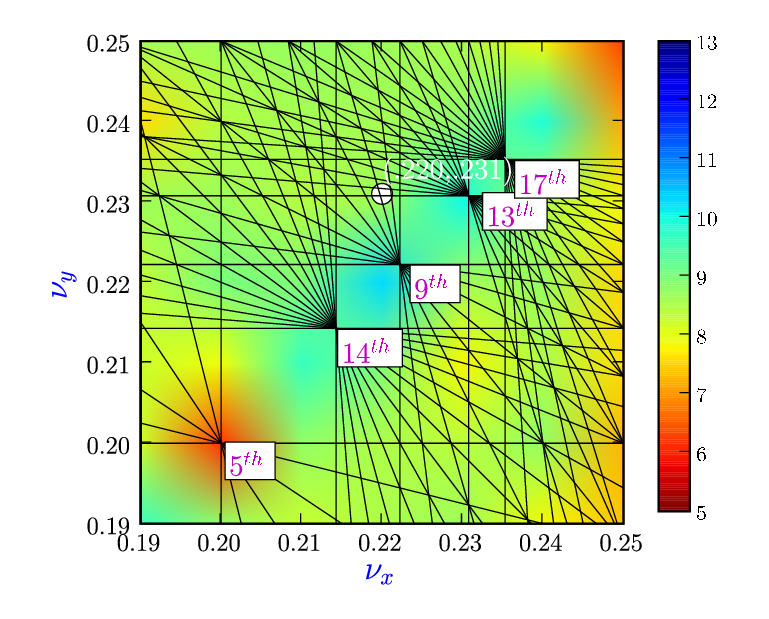}}
\par\end{centering}

\caption{Tune scan of dynamic aperture at gold collision energy: (a) $d_{y}=8$
$\sigma$ and (b) $d_{y}=6$ $\sigma$. Working point $\left(28.220,29.231\right)$
is plotted as a white dot. Red indicates low dynamic apertures while
blue indicates high dynamic apertures. The unit of dynamic aperture
is $\sigma$. \label{fig:da-ts-gc}}

\end{figure}

\begin{figure}
\begin{centering}
\subfloat[]{

\centering{}\includegraphics[bb=135bp 250bp 470bp 530bp,clip,scale=0.5]{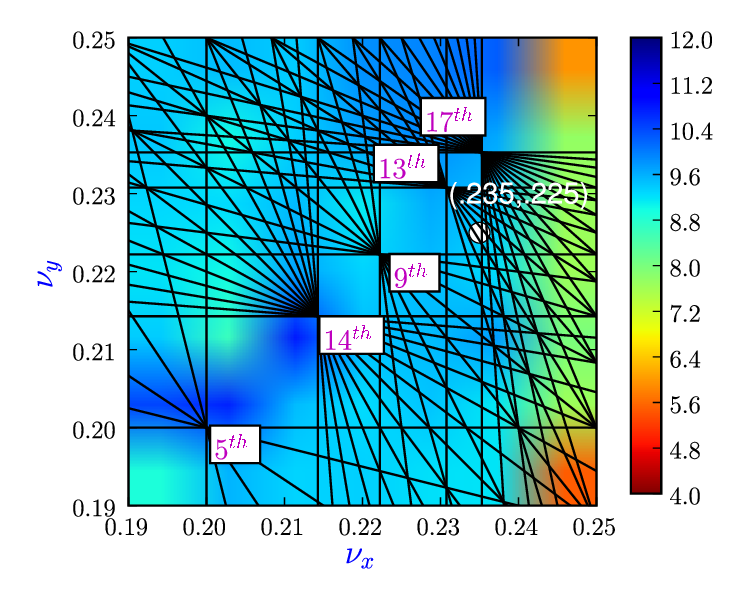}}
\par\end{centering}

\begin{centering}
\subfloat[]{

\centering{}\includegraphics[bb=135bp 250bp 470bp 530bp,clip,scale=0.5]{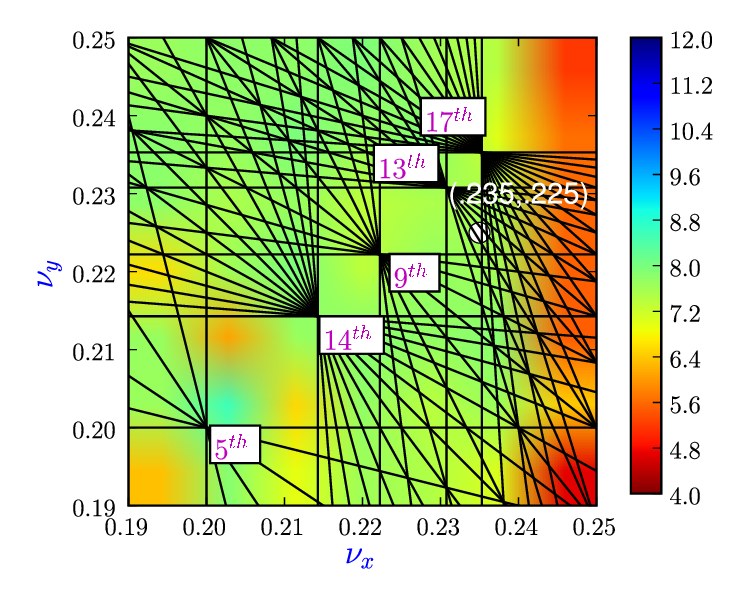}}
\par\end{centering}

\caption{Tune scan of dynamic aperture at deuteron collision energy: (a) $d_{y}=8$
$\sigma$ and (b) $d_{y}=6$ $\sigma$. Working point $\left(28.235,29.225\right)$
is plotted as a white dot. Red indicates low dynamic apertures while
blue indicates high dynamic apertures. The unit of dynamic aperture
is $\sigma$. \label{fig:da-ts-dau}}

\end{figure}

Figure \ref{fig:da-ts-gi} shows the contour plots of dynamic apertures
over transverse tunes for two different wire separations for gold
injection energy. Red indicates low dynamic apertures around 3 $\sigma$
while blue indicates high dynamic apertures around 11 $\sigma$. %
 It is found that at all wire separations, the largest dynamic apertures
are distributed along a band parallel to the diagonal, i.e., $\nu_{x}-\nu_{y}\simeq0.02$.
On the other hand, the zone along $\nu_{x}-\nu_{y}\simeq0.03$ has
the smallest dynamic apertures at all separations. This scan indicates
that the nominal tune $\left(28.230,29.216\right)$ is close to optimal.
Furthermore, a sharper drop in dynamic aperture is observed near the
$5^{th}$ resonances than at other resonances as the separation decreases
from 8 $\sigma$ to 4 $\sigma$. Figure \ref{fig:da-ts-gc} shows
the dynamic apertures over transverse tunes for gold collision energy.
Here red stands for 5 $\sigma$ while blue for 11 $\sigma$. %
 Since the IR multipoles cause a large drop in dynamic aperture at
collision energy and the wire has a relatively smaller effect, we
exclude the IR multipole errors in order to see the effect of the
wire alone. As expected, the smallest dynamic aperture is observed
near $4^{th}$ and $5^{th}$ resonances. At all wire separations,
the largest dynamic apertures are distributed nearly along the diagonal
between $\nu_{x}=0.21$ and $\nu_{x}=0.24$. The effect of the wire
on the dynamic aperture of the deuteron beam is presented in Fig.
\ref{fig:da-ts-dau}. %
 It is clear that the reduction of the dynamic aperture is dominant
near $4^{th}$ resonance. A notable variation is seen near a circular
band, i.e., $\nu_{x}^{2}+\nu_{y}^{2}\simeq0.21^{2}$, when the beam-wire
separation is small, while the circular band is not detected in the
gold beam. Indeed, the large stability boundary is distributed all
over the tune space except the particular bands. Since the wire kick
at a large separation is equivalent to long-range beam-beam interaction,
these scans can be interpreted as showing the effects of a long-range
interaction at different tunes.

\subsection{Beam Transfer Functions}

The beam transfer function (BTF) is defined as the beam response to
a small external longitudinal or transverse excitation at a given
frequency. BTF diagnostics are widely employed in modern storage rings
due to its non-destructive nature. The beam response is observed usually
in a downstream pickup while either a stripline kicker or RF cavity
excites the betatron or synchrotron oscillation over the tune spectrum.
The fundamental applications of BTF are to measure the transverse
tune and tune distribution by exciting betatron oscillation, to analyze
the beam stability limits, and to determine the impedance characteristics
of the chamber wall and feedback system \cite{Borer,Chou,Kornilov}.
Since this is one of the observables in RHIC, we calculate the BTF
to benchmark another output from the code against measurements.

\begin{figure*}
\begin{centering}
\subfloat[]{

\centering{}\includegraphics[scale=0.45]{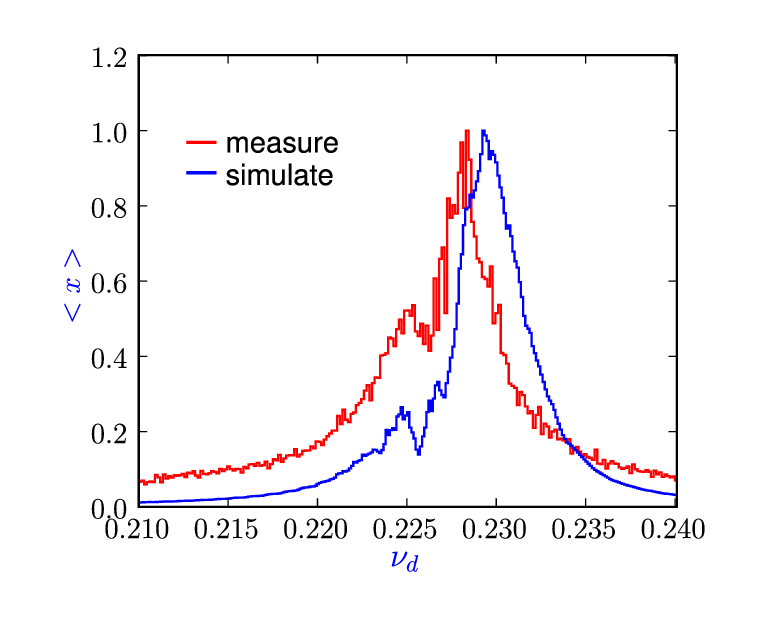}}\subfloat[]{\centering{}\includegraphics[scale=0.45]{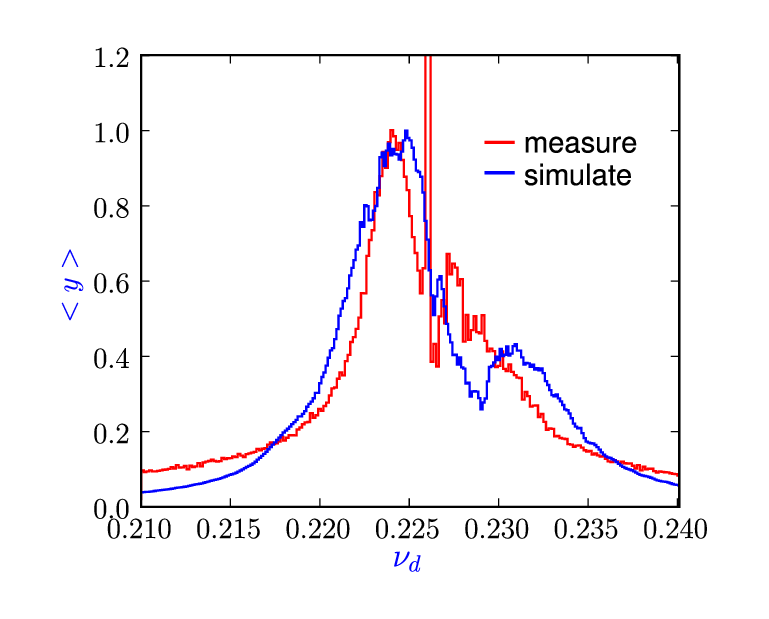}}
\par\end{centering}

\begin{centering}
\subfloat[]{\centering{}\includegraphics[scale=0.45]{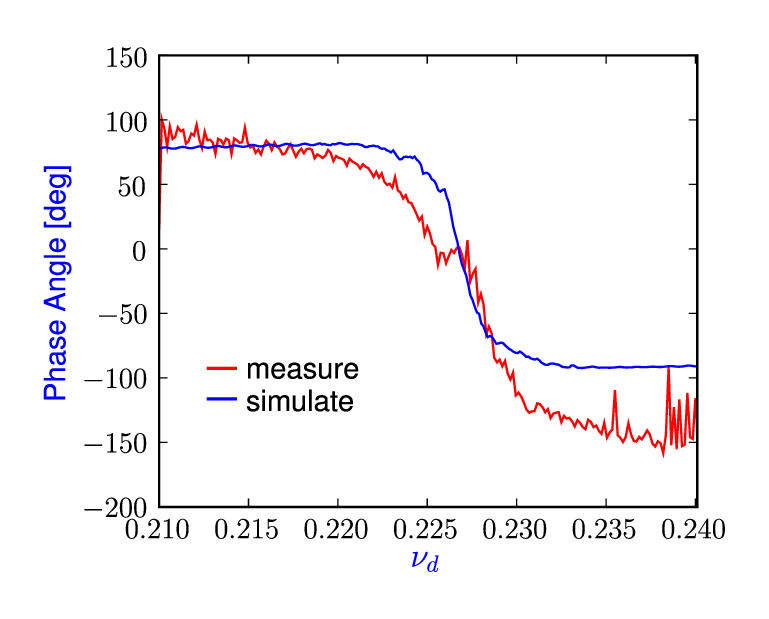}}\subfloat[]{

\centering{}\includegraphics[scale=0.45]{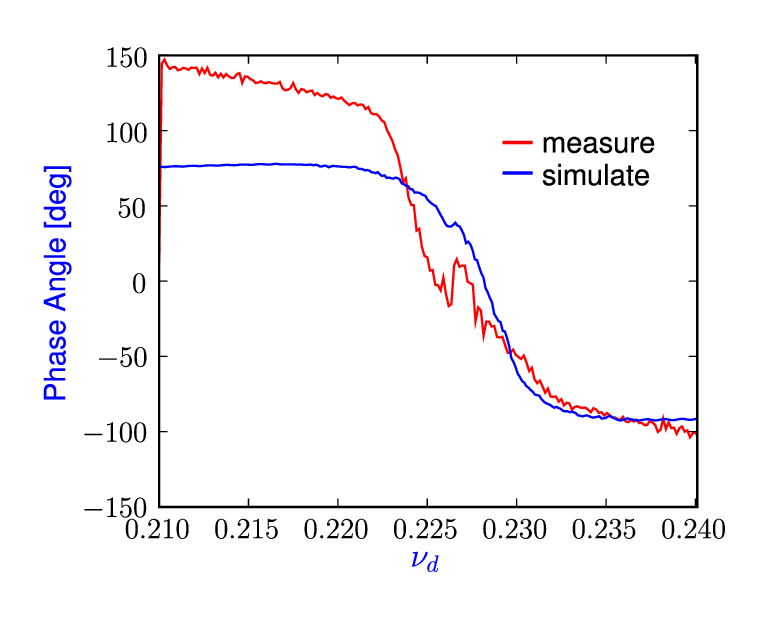}}
\par\end{centering}

\caption{Amplitude of (a) horizontal and (b) vertical beam transfer functions,
and phase angle of (c) horizontal and (d) vertical BTF for deuteron
beam. The amplitude is in arbitrary units, and the phase angle is
in degree. The wire strength is 125 Am. \label{fig:dau_btf_nw}}

\end{figure*}

Figure \ref{fig:dau_btf_nw} presents the beam transfer functions
for deuteron collision energy and nominal tune $\left(28.228,29.225\right)$
obtained by applying a sinusoidal driving force to a beam in transverse
plane and tracking the excited particles over 1024 turns at each excitation
frequency of the kicker. %
 The driving frequency is swept from $\left(n+0.21\right)f_{rev}$
to $\left(n+0.24\right)f_{rev}$ in steps of $2\times10^{-3}f_{rev}$,
where $n$ is the integer tune and $f_{rev}$ the revolution frequency
which are listed in Table \ref{tab:rhic}. Through this frequency
scan, we compute the amplitude of beam response and its phase. However,
as the excitation strength is increased, it is observed that the response
curves are different according to the direction of the frequency sweep,
i.e., downward or upward \cite{Minty,Byrd}. The driving amplitude
should be applied as low as the response can be detected. In the simulations,
the driving amplitude is chosen as $10^{-4}$ $\sigma_{x^{'},y^{'}}$,
where $\sigma_{x^{'},y^{'}}$ are the transverse rms slopes. Besides,
the response profile is affected by the sweeping rate. A relaxation
time needs therefore to be applied just before starting the evaluation
of beam transfer function at each successive driving frequency, because
the directional dependence of the response amplitude may be due to
the remnants of nonlinear oscillations driven by former excitations.
In order to avoid the uncertainty, we reload fresh particles with
distribution identical to the initial one.

The response curve reveals two peaks, as shown in Fig. \ref{fig:dau_btf_nw}
(a) and (b): One peak is close to 0.230 which is the horizontal tune,
and the other is 0.225 which is the vertical tune. The amplitude is
in arbitrary units and normalized by the peak amplitude. We observe
signs of transverse coupling in these BTFs, the vertical tune shows
up as a lower peak in the horizontal BTF and similarly the horizontal
tune appears in the vertical BTF. It is clear that due to resonance
with the external force, the betatron amplitude of particles grows
when the driving frequency is close to the betatron frequency. However,
we observe valleys in the response at $\nu_{d}=0.226$ and $\nu_{d}=0.228$
for both horizontal and vertical amplitudes. They are located at the
resonance line of order 9, and may stem from the detuning of particles
away from the resonance. For a sinusoidal driving force, the phase
of the beam transfer function change its sign through the resonance
frequency \cite{Chao}. Since the simulation does not account for
the phase change due to cable delays between the pickup and the network
analyzer, the absolute phases will not agree between simulations and
measurements. We can however expect a similar phase variation across
the resonance. Both simulation and measurement results show the flip
of the phase angle as shown in Fig. \ref{fig:dau_btf_nw} (c) and
(d). The small differences in the zero crossing phase and the differences
in peak locations of the amplitudes between simulation and measurement
are likely due to the external impedances and the nonlinearities not
included in the simulation.

\begin{figure*}
\begin{centering}
\subfloat[]{

\centering{}\includegraphics[scale=0.45]{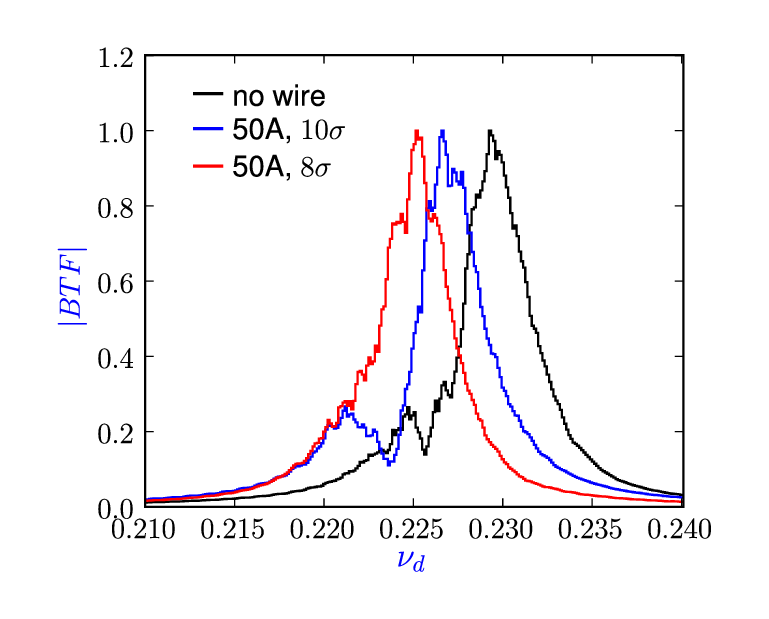}}\subfloat[]{\centering{}\includegraphics[scale=0.45]{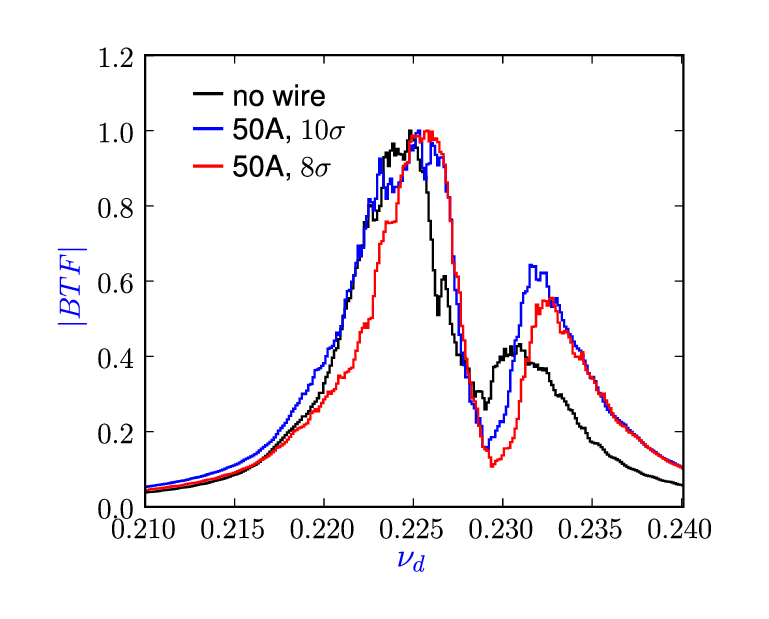}}
\par\end{centering}

\begin{centering}
\subfloat[]{\centering{}\includegraphics[scale=0.45]{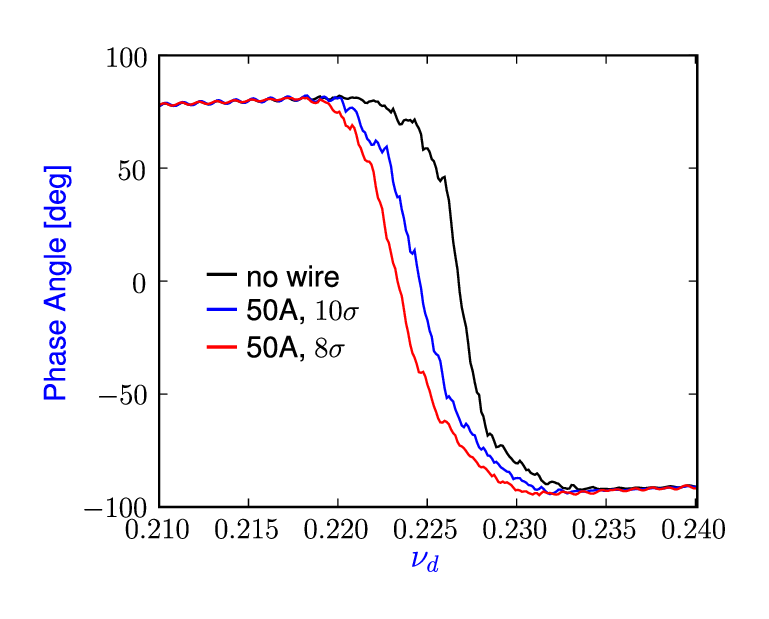}}\subfloat[]{

\centering{}\includegraphics[scale=0.45]{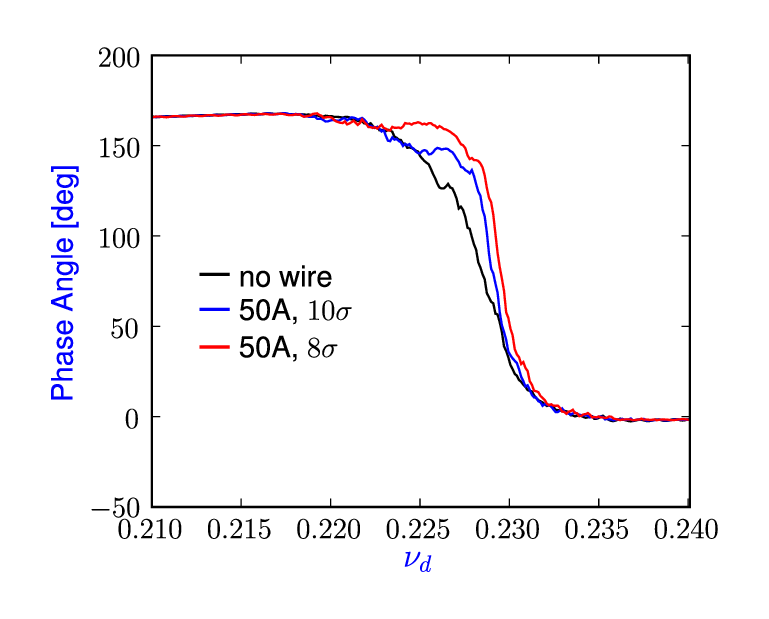}}
\par\end{centering}

\caption{Amplitude of (a) horizontal and (b) vertical beam transfer functions,
and phase angle of (c) horizontal and (d) vertical BTF for deuteron
beam. The amplitude is in arbitrary units, and the phase angle is
in degree. The wire current is set 50A. The separation is in unit
of rms beam size at wire location.\label{fig:dau-btf_w}}

\end{figure*}

The influence of the current carrying wire to the beam transfer function
is presented through the comparison of Fig. \ref{fig:dau-btf_w} which
shows the beam amplitude and phase response obtained without the wire
and with the wire powered at 50A and at two separations in a deuteron
beam study.%
 The shift of a peak location of the amplitude increases as the wire
separation decreases while the width of the amplitude response widens,
as shown in Fig. \ref{fig:dau-btf_w} (a) and as expected from the
increase in the size of the tune spread. The wire with 10 $\sigma$
and 8 $\sigma$ separations shifts the peak location of the horizontal
response by $3.4\times10^{-3}$ and $4.1\times10^{-3}$ respectively,
which are equivalent to the tune shift of zero amplitude particles.
As shown in Fig. \ref{fig:dau-btf_w} (c), the wire moves the phase
profile leftward commensurately with the shift of the amplitude. The
BTF response contains a wealth of information about the beam, a detailed
analysis of measured and simulated BTFs is left to a subsequent publication.

\subsection{Diffusion Coefficients}

We have calculated the beam diffusion due to non-linear particle dynamics
which includes the nonlinearities from the machine itself, the head-on
beam-beam interactions, and the current carrying wire. Growth of particle
amplitudes may be described by a diffusion in action variables. The
diffusion coefficients can characterize the effects of the nonlinearities
present in the accelerator, and can be used to find numerical solutions
of a diffusion equation \cite{hjkim}. The solutions yield the time
evolution of the beam density distribution function for a given set
of machine and beam parameters. This technique enables us to follow
the beam intensity and emittance growth for the duration of a luminosity
store, something that is not feasible with direct particle tracking.
Here we will focus only the calculation of the diffusion coefficients
and compare them with past measurements in RHIC.

The transverse diffusion coefficients can be calculated numerically
from \begin{equation}
\begin{aligned}D_{ij}\left(a_{i},a_{j}\right) & =\frac{1}{N}\left\langle \left(J_{i}(a_{i},N)-J_{i}(a_{i},0)\right)\right.\\
 & \qquad\quad\left.\left(J_{j}(a_{j},N)-J_{j}(a_{j},0)\right)\right\rangle ,\end{aligned}
\label{eq:rd-4}\end{equation}
 where $J_{i}\left(a_{i},0\right)$ is the initial action at an amplitude
$a_{i}$, $J_{i}\left(a_{i},N\right)$ the action at an amplitude
$a_{i}$ after $N$ turns, $\left\langle \right\rangle $ the average
over simulation particles, and $(i,j)$ are the horizontal $x$ or
the vertical $y$ coordinates. Equation \prettyref{eq:rd-4} is averaged
over a certain number of turns to eliminate the fluctuation in action
due to the phase space structure, e.g. resonance islands. In the simulations,
the tracking code evaluates the diffusion coefficients in two-dimensional
action space with the boundary determined by the dynamic aperture
obtained in the previous section. We load the initial particle distribution
which is built by using 100 particles placed at the same transverse
action. The tracking is performed for $10^{6}$ turns. The diffusion
coefficient is averaged every $10^{4}$ turns at which the coefficient
approaches approximately the asymptotic limit. The above process is
performed over the transverse action space with the boundary typically
set at an amplitude of 7 $\sigma$. 

\begin{figure}
\begin{centering}
\subfloat[]{\begin{centering}
\includegraphics[scale=0.5]{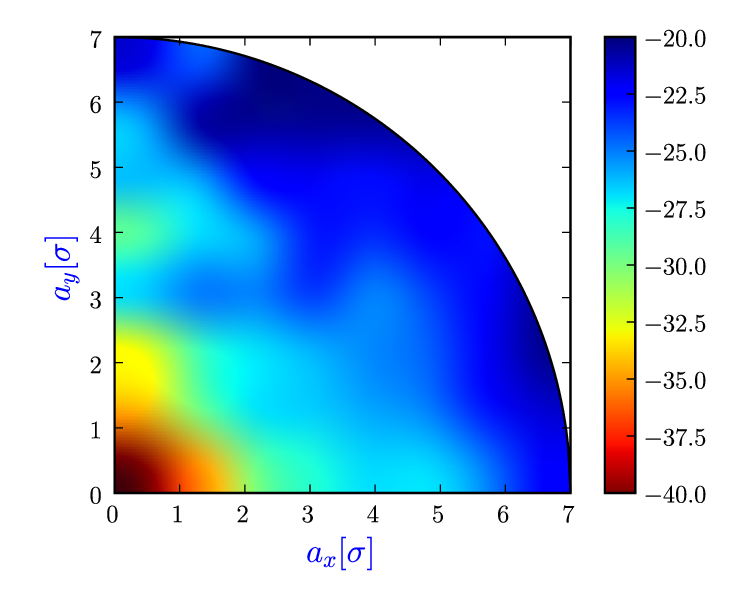}
\par\end{centering}

}
\par\end{centering}

\begin{centering}
\subfloat[]{\begin{centering}
\includegraphics[scale=0.5]{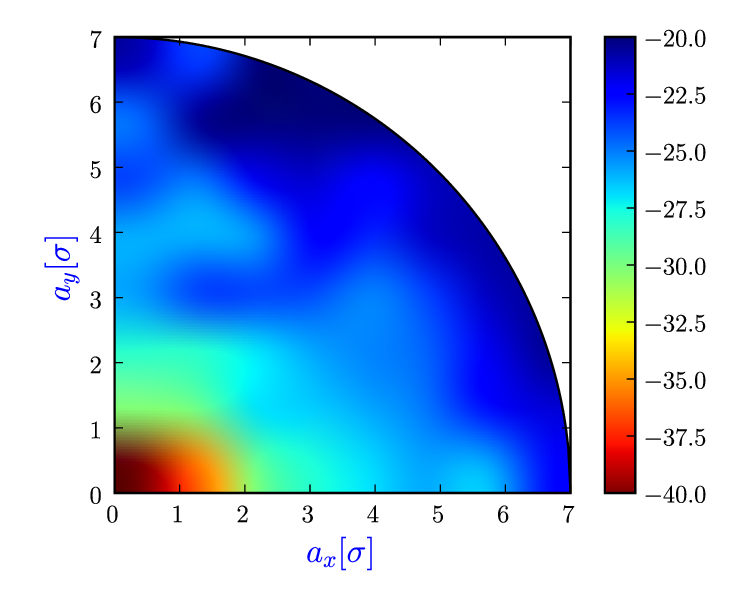}
\par\end{centering}

}
\par\end{centering}

\caption{Contour plot of (a) the horizontal diffusion coefficient $D_{xx}$
and (b) the vertical diffusion coefficient $D_{yy}$ for gold beam
at collision energy. The coefficients are calculated at the Blue ring.
The color assignment is logarithmically scaled in the plots. \label{fig:cont_dc}}

\end{figure}

Figure \ref{fig:cont_dc} presents the contour plot of the horizontal
and vertical components of the diffusion coefficients for the gold
beam at collision energy. %
 In these plots we observe that the diffusion coefficients have angular
dependence. The horizontal diffusion $D_{xx}$ depends strongly upon
the horizontal action $J_{x}$ while its dependence on the vertical
action is weak. Similarly the vertical diffusion $D_{yy}$ depends
primarily on the vertical action. The diffusion coefficients scale
exponentially with initial actions, specially at small actions, in
both horizontal and vertical directions but at very different rates
in each direction. A single exponential fit does not however suffice
but a combination of at least two different exponentials is required
to describe the growth of the diffusion coefficients from the origin
to the dynamic aperture. 

\begin{figure}
\begin{centering}
\subfloat[]{

\centering{}\includegraphics[scale=0.45]{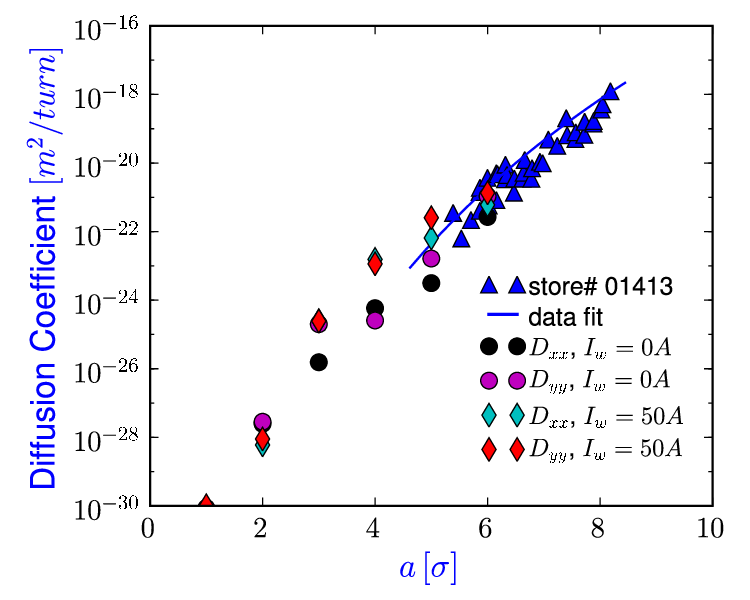}}
\par\end{centering}

\begin{centering}
\subfloat[]{

\centering{}\includegraphics[scale=0.45]{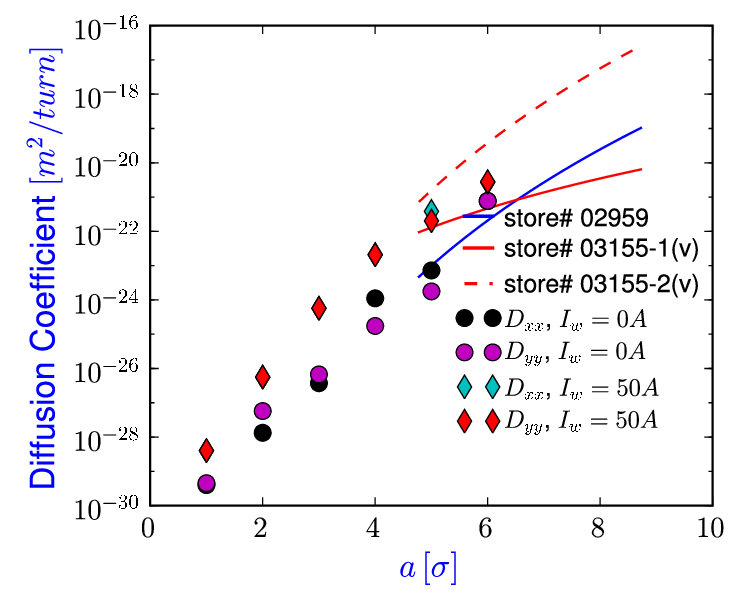}}
\par\end{centering}

\begin{centering}
\subfloat[]{

\centering{}\includegraphics[scale=0.45]{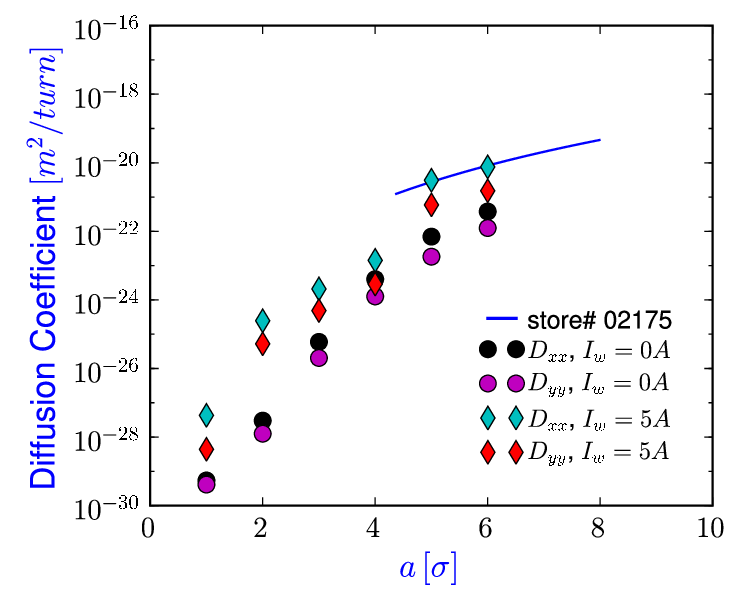}} 
\par\end{centering}

\caption{Plot of diffusion coefficients of (a) gold, (b) deuteron, and (c)
proton stores of RHIC. The coefficients are calculated at the Blue
ring. The wire-beam separation distance is 8 $\sigma$ at a wire location.
The coefficients were measured and fitted in stores in previous years
\cite{Fliller}. \label{fig:diff_coeff}}

\end{figure}

Figure \ref{fig:diff_coeff} shows the diffusion coefficients for
three RHIC situations: gold beam at store, deuteron beam at store,
and proton beam at store. The coefficients are plotted at the action
$J=\sqrt{J_{x}^{2}+J_{y}^{2}}$ after averaging them at the same action
and are compared with the measurements obtained by fitting the time-dependent
loss rate after moving a collimator into and out from the beam \cite{Fliller}.
The loss rate due to the movement of the collimator can be described
by \cite{Seidal} \begin{equation}
\begin{aligned}\dot{N}^{\left(1\right)}\left(t\right) & =a_{0}\left(1+\frac{\Delta z}{\sqrt{\pi R}\left(t-t_{0}\right)}\right)+a_{1},\\
\dot{N}^{\left(2\right)}\left(t\right) & =a_{0}\text{erfc}\left(\frac{\Delta z}{\sqrt{\pi R}\left(t-t_{0}\right)}\right)+a_{1},\end{aligned}
\label{eq:rd-5}\end{equation}
where the superscript (1) and (2) stands for the inward and outward
movement of the collimator, $\Delta z$ the change in z due to the
collimator movement, $R$ the diffusion coefficient, and $a_{0}$
and $a_{1}$ are constants. The diffusion coefficient $R$ can be
obtained directly from the fit of measured loss rates. In Fig. \ref{fig:diff_coeff},
the vertical axis is a logarithmic scale. It should be noted that
dependence of diffusion coefficients on the initial action is exponential
at small amplitudes and power law like at larger amplitudes. However,
since the measured coefficients are fitted by a power law, i.e. $D\sim J^{n}$,
they agree with simulations only at large actions. Since the collimators
were not moved into the beam core, the diffusion coefficients were
not measured at small actions. Conversely it is difficult to calculate
the coefficients in simulations at large action because some of the
particles are lost quickly. The effects of wire on diffusion coefficients
are considerable at all action amplitudes, as shown in Fig. \ref{fig:diff_coeff}.
For example, the diffusion coefficient at 3 $\sigma$ amplitude becomes
20 times larger when the wire with strength 125 Am and separation
$d_{y}=8$ $\sigma$ is applied. First we observe that the relative
increase of diffusion coefficients at below 3 $\sigma$ amplitude
for the deuteron beam is higher than that for the gold beam. In general
the diffusion process depends on the particle motion in phase space
and the resonance structure due to the nonlinearities. The differences
in tunes change the cross-talk between the different nonlinearities
as well as the resonance driving terms which are likely to be responsible
for the differences in diffusion. The enhanced diffusion at near 3
$\sigma$ amplitude for the deuteron beam leads to significant increase
of particle loss under the simulation conditions (see the following
section).  %

\subsection{Beam losses as a function of wire-beam separation}

The beam lifetime in RHIC is determined by the inelastic interactions
and beam-beam interactions when the beams are in collision as well
as nonlinearities of machine elements, intra-beam scattering and residual
gas scattering. In this section we will focus on the impact of the
wire on the beam loss rates as the beam-wire separation is changed.
In this study, the initial beam distribution is a hollow Gaussian
in transverse phase space and a normal Gaussian in longitudinal phase
space. The initial transverse beam sizes are obtained from the RHIC
optics and typical initial emittances while the bunch length is taken
from measured values. 

\begin{figure}
\begin{centering}
\includegraphics[bb=140bp 255bp 465bp 525bp,clip,scale=0.5]{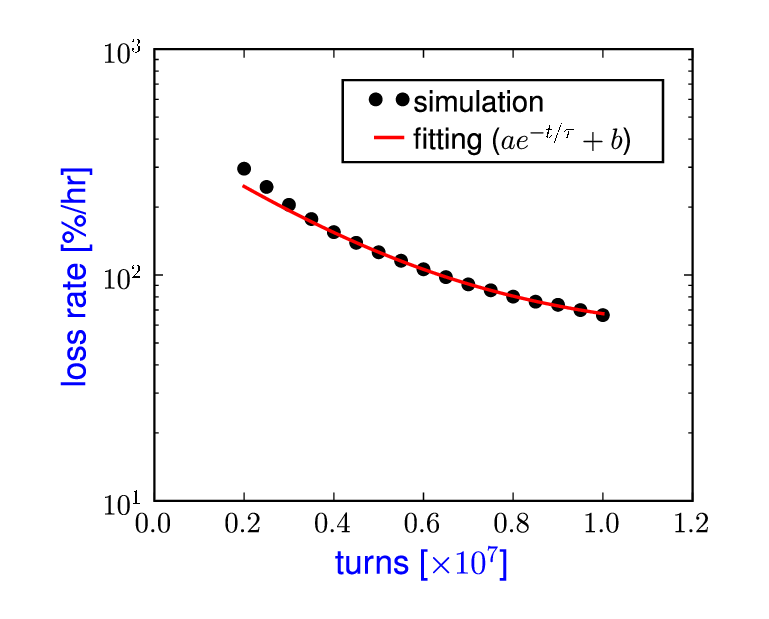}
\par\end{centering}

\caption{Plot of the variation of loss rate versus the number of tracking turns.
The loss rate is estimated from the asymptotic value of the exponential
decay fit, i.e., $ae^{-t/\tau}+b$. \label{fig:loss-fit}}

\end{figure}

The tracking is done with $5\times10^{3}$ macroparticles, and carried
out over $10^{7}$ turns for each wire separation. The loss rates
are estimated from the asymptotic limit by extrapolating the simulated
loss rate from $10^{7}$ turns to infinity as shown in Fig. \ref{fig:loss-fit}.
 It is observed that in the beginning of the simulation, the loss
rate decreases exponentially rapidly and then approaches a constant
rate at later times. We apply, therefore, an exponential decay fit,
i.e., $ae^{-t/\tau}+b$, to the loss rate data between $4\times10^{6}$
and $1\times10^{7}$ turns. The asymptotic loss rate is the parameter
$b$.

\begin{figure}
\begin{centering}
\subfloat[]{

\centering{}\includegraphics[bb=130bp 250bp 465bp 525bp,clip,scale=0.475]{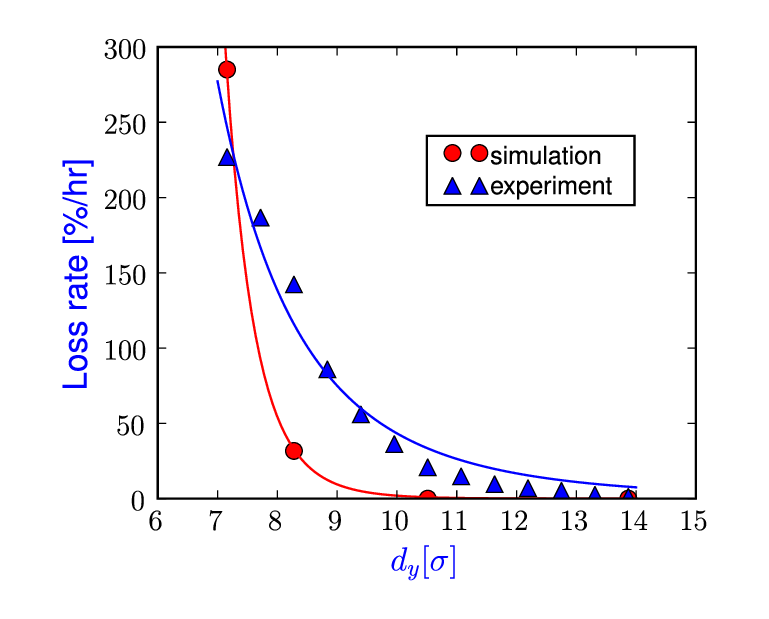}}
\par\end{centering}

\begin{centering}
\subfloat[]{

\centering{}\includegraphics[bb=130bp 250bp 465bp 525bp,clip,scale=0.475]{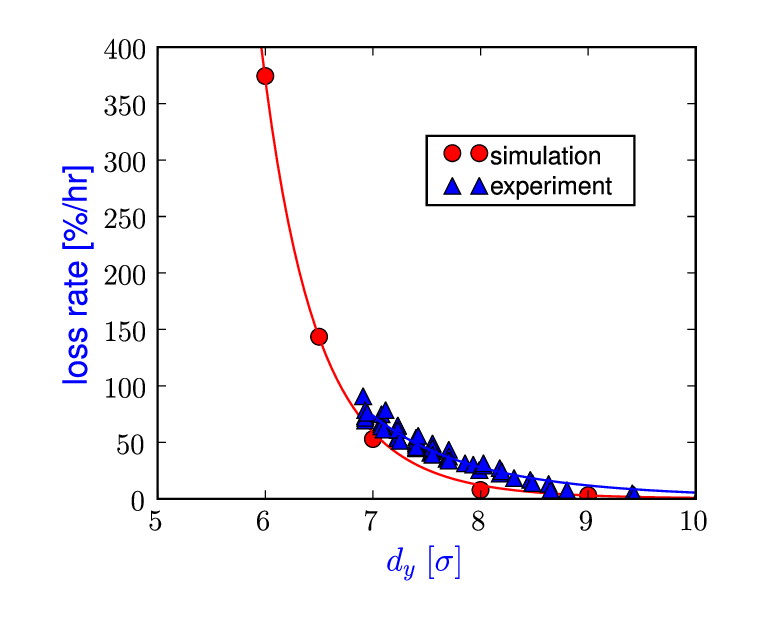}}
\par\end{centering}

\begin{centering}
\subfloat[]{

\centering{}\includegraphics[bb=130bp 250bp 465bp 525bp,clip,scale=0.475]{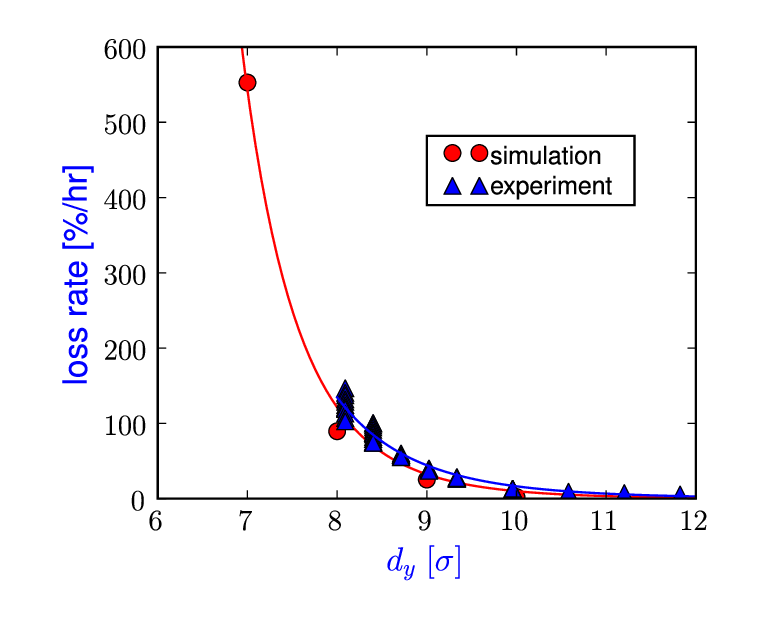}}
\par\end{centering}

\caption{Comparison of the simulated beam loss rates with the measured as a
function of separations. (a) gold beam at injection energy, (b) gold
beam at collision energy, (c) deuteron beam at collision energy. Wire
strength is 125 Am. \label{fig:loss-rate}}

\end{figure}

Figure \ref{fig:loss-rate} plots the asymptotic beam loss rate due
to the wire as a function of beam-wire separation distance for the
case of Table \ref{tab:rhic}. %
 For the gold injection energy study, we include the chromaticity
correcting sextupoles, chromaticity set to two units, and the wire.
It can be seen in Fig. \ref{fig:dynamic} (a) that the electromagnetic
force of the wire decreases the dynamic aperture significantly compared
to the case when the wire is not present. The particle loss rate of
the beam shows a sharp increase wire separations smaller than 8 $\sigma$.

For the collision energy simulations, we include the nonlinear field
errors in the triplets, the head-on beam-beam interactions, and the
nonlinearities in the injection energy study. The onset of beam losses,
seen in Fig. \ref{fig:loss-rate} (b) and (c), is observed at 8 $\sigma$
and 9 $\sigma$ for gold and deuteron beams respectively. In all three
cases, the threshold separation for the onset of sharp losses observed
in the measurements and simulations agree to better than 1 $\sigma$.
It is also significant that the simulated loss rates at 7 and 8 $\sigma$
separation for the gold beam and 8 and 9 $\sigma$ for the deuteron
beam are very close to the measured loss rates. At fixed separation,
the wire causes a much higher beam loss with the deuteron beam than
with the gold beam. For example, the loss-rate for the gold beam at
a 8 $\sigma$ separation is about 10 \%/hr while for the deuteron
beam the loss rate is about an order of magnitude higher both in measurements
and simulation. This difference is not reflected in the dynamic apertures,
shown in Figure \ref{fig:dynamic}, which are about the same for the
two cases at the same beam-wire separation. Nor is this difference
correlated to the tune footprint and the resonance lines, seen in
Fig. \ref{fig:footprint}, where the footprint for the deuteron beam
is free of resonance lines lower than the $12^{th}$ while the gold
footprint is spanned by the $9^{th}$ and $12^{th}$ order resonances.
However the frequency diffusion maps with the wire, seen in Fig. \ref{fig:fd_au}
and Fig. \ref{fig:fd_dau}, show greater diffusion in the deuteron
case than in the gold case. This correlation of the frequency diffusion
maps with the loss rates observed here deserves to be studied more
deeply and tested for validity in other accelerators. The action diffusion
seen in Fig. \ref{fig:diff_coeff} is also larger in the deuteron
beam than the gold beam at small amplitudes by one to two orders of
magnitude. Thus both frequency and action diffusion seem to be better
correlated with loss rates than the traditional short term indicators
like footprints and dynamic aperture.

Changing the beam-wire separation changes several parameters including
the tunes, the tune spread, the resonance driving terms etc. The wire
separation alone cannot describe the change in the dynamics that influences
the loss rates. Nonetheless the loss rates seen in Fig. \ref{fig:loss-rate}
can be fitted to a power law in the separation. The results for the
three cases for the measurements and simulations are shown in Table
\ref{tab:loss-coeff}. %
\begin{table}
\begin{centering}
\begin{tabular}{cccc}
\hline 
 & \multicolumn{2}{c}{gold beam} & \multirow{2}{2cm}{deuteron beam}\tabularnewline
 & injection & store & \tabularnewline
\hline
Measurement & -5.2 & -7.3 & -9.4\tabularnewline
Simulation & -14.2 & -11.9 & -11.2\tabularnewline
\hline
\end{tabular}
\par\end{centering}

\caption{Dependence of loss rate on the beam-wire separation: The loss rates
are fitted to $\tau\propto d_{y}^{\alpha}$, where $d_{y}$ is the
separation, and the power $\alpha$ is listed on the table.\label{tab:loss-coeff}}

\end{table}
The simulations have studied smaller separations than would be practical
in RHIC - the large loss rates seen at the smallest separation would
have quenched the machine. However, the simulations show the steep
climb in the loss rate beyond a threshold separation and consequently
will have a higher power law behavior. The power law at injection
is somewhat higher than at collision but the power laws for gold and
deuteron beams are very close despite the large difference in loss
rates. We expect that the power laws depends on the details of the
machine and not to be universal. In fact different power laws have
been reported for the SPS and the Tevatron \cite{Lebrun,Zimmermann,Dorda-2}.

\section{Summary\label{sec:summary}}

In order to study the effects of the machine nonlinearities and the
beam-beam interactions, including strong localized long-range beam-beam
interactions, we have developed a six-dimensional weak-strong code
\texttt{bbsimc}. Machine nonlinearities, beam-beam collisions and
the field of a current carrying wire can be included in the model.

We have studied the effects of the wire on the beam dynamics in three
different cases: gold beam at injection, gold beam at store, and deuteron
beam at store.  Results show that the betatron tune change due to
the wire is well tracked by the simulation, and that the stability
of particle motion is strongly influenced by the wire which causes
a significant increase in tune spread and diffusion for both gold
and deuteron beam. Tune diffusion with the gold beam and the wire
is appreciable only at amplitudes larger than 2 $\sigma$ while the
tune diffusion with the deuteron beam and the wire is significantly
larger extending to the beam core at amplitudes down to 0.5 $\sigma$.

At injection energy, the dynamic aperture is largely determined by
the wire, while the wire and the IR multipoles have a major impact
on the stability boundary at collision. With the beam-wire separation
only along the vertical axis, it is found that the stability boundary
near the vertical axis is linearly proportional to the beam-wire separation.
We observe, from the tune scan of dynamic aperture, that at injection
the largest and the smallest dynamic apertures are distributed along
a band parallel to the zone along $\nu_{x}-\nu_{y}\simeq0.02$ and
$\nu_{x}-\nu_{y}\simeq0.03$ respectively at all wire separations
while at collision the largest dynamic apertures are distributed nearly
along the diagonal between $\nu_{x}=0.21$ and $\nu_{x}=0.24$. By
modeling the stripline kicker we obtain the amplitude and phase angle
of the transverse BTF. Results show that the betatron tune is well
identified by the simulation, and that the wire changes characteristics
of the beam response significantly in the horizontal plane, similar
to the result for the tune shift.

In simulations the wire enhances the diffusion. This effects is particularly
pronounced in the plane of the wire, and at amplitudes of about 3
$\sigma$, where the wire can increase the diffusion rate by more
than an order of magnitude. The action diffusion for the deuteron
beam is larger than for the gold beam, similar to the result for frequency
diffusion. Simulations of the beam loss rate when the wire is present
are in good agreement with the experimental observations. The threshold
separation at which there is a sharp rise in the loss rates agree
to better than 1 $\sigma$. The loss rates with the deuteron beam
are found in measurements and simulations to be nearly an order of
magnitude higher than with the gold beam. Comparisons of tune footprints
and dynamic apertures of these two cases show no indication that the
loss rates may be higher with the deuteron beam and in fact the tune
footprint with the deuteron beam is free of low order resonances.
However frequency diffusion and action diffusion with the deuteron
beam are substantially higher than with the gold beam implying that
these diffusion measures may be better indicators of loss rates. 

In this paper simulation results of the beam-wire interactions were
compared with measurements taken when the long-range beam-beam interactions
were not present. It is expected that compensation of parasitic interactions
using the wires will be tested during the RHIC run in 2009. We plan
to compare simulations with these forthcoming experimental results
and also to determine the effectiveness of the compensation of long-range
interactions in the LHC.
\begin{acknowledgments}
We thank Y. Luo and R. Calaga for help with the RHIC lattice. V. Boocha,
B. Erdelyi and V. Ranjbar made significant contributions to the development
of the code. Some of the parallel computations were performed at the
NERSC facility at LBL. This work was supported by the US-LARP collaboration
which is funded by the US Department of Energy. \end{acknowledgments}

\end{document}